\pdfoutput=1
\documentclass[12pt,a4paper]{article}
\usepackage{dcolumn,booktabs}
\newcolumntype{d}[1]{D{.}{.}{#1}}
\newcommand\mc[1]{\multicolumn{1}{c}{#1}}

\usepackage{float}
\usepackage{ifthen} 
\newboolean{pdflatex}
\setboolean{pdflatex}{true} 

\newboolean{articletitles}
\setboolean{articletitles}{true} 

\newboolean{uprightparticles}
\setboolean{uprightparticles}{false} 

\usepackage{multirow}
\usepackage{booktabs}
\usepackage{subcaption}
\usepackage{wrapfig}
\usepackage{bm}

\newcommand{\ResultRelaBRBsDsoKDstK}{\ensuremath{0.409\pm0.019\pm0.020\pm0.010}}

\newcommand{\ResultBRBsDsoKDstK}{\ensuremath{(2.49\pm0.11\pm0.12\pm0.25\pm0.06)\times10^{-5}}}

\newcommand{\ResultRelaBRBzDsoKDstK}{\ensuremath{0.084\pm0.003\pm0.006}}

\newcommand{\ResultBRBzDsoKDstK}{\ensuremath{(0.510\pm0.021\pm0.036\pm0.050)\times 10^{-5}}}

\newcommand{\ResultK}{\ensuremath{1.89\pm0.24\pm 0.06}}
\newcommand{\ResultPsi}{\ensuremath{1.81\pm0.20\pm 0.11}}

\newcommand{\ResultKNoSys}{\ensuremath{1.89\pm0.24}}
\newcommand{\ResultPsiNoSys}{\ensuremath{0.58\pm0.06}}

\newcommand{\ResultA}{\ensuremath{1.11\pm0.15\pm 0.06}}
\newcommand{\ResultB}{\ensuremath{0.70\pm0.09\pm 0.04}}

\def\calR {\ensuremath{{\mathcal{R}}}\xspace}
\def\calN {\ensuremath{{\mathcal{N}}}\xspace}

\def\Dsop {{\ensuremath{\D_{\squark1}(2536)^{+}}}\xspace}
\def\Dsom {{\ensuremath{\D_{\squark1}(2536)^{-}}}\xspace}

\def\Dsomp {{\ensuremath{\D_{\squark1}(2536)^{\mp}}}\xspace}
\def\Dso {{\ensuremath{\D_{\squark1}}}\xspace}
\newcommand{\BdsDsoK}{\decay{B_{(s)}^0}{\Dsomp\Kpm}}
\newcommand{\BdsDsomK}{\decay{B_{(s)}^0}{\Dsom\Kp}}

\newcommand{\BsDsoK}{\decay{\Bs}{\Dsomp\Kpm}}

\newcommand{\BdDKK}{\decay{\Bz}{\Dzb\Kp\Km}}

\def\paperauthors{LHCb collaboration} 
\def\paperasciititle{Observation of the B_(s)0 to D_s1(2536)K decays} 
\def\papertitle{Observation of the decays \BdsDsoK } 
\def\paperkeywords{{High Energy Physics}, {LHCb}} 
\def\papercopyright{\the\year\ CERN for the benefit of the LHCb collaboration} 
\def\paperlicence{CC BY 4.0 licence}
\def\paperlicenceurl{https://creativecommons.org/licenses/by/4.0/}

\usepackage[top=1in, bottom=1.25in, left=1in, right=1in]{geometry}

\usepackage{microtype}
\usepackage{lineno}  
\usepackage{xspace} 
\usepackage{caption} 

\usepackage{graphicx}  
\usepackage{color}
\usepackage{colortbl}
\graphicspath{{./figs/}} 

\usepackage{amsmath} 
\usepackage{amssymb}
\usepackage{amsfonts}
\usepackage{upgreek} 

\newcommand*\patchAmsMathEnvironmentForLineno[1]{%
\expandafter\let\csname old#1\expandafter\endcsname\csname #1\endcsname
\expandafter\let\csname oldend#1\expandafter\endcsname\csname
end#1\endcsname
 \renewenvironment{#1}%
   {\linenomath\csname old#1\endcsname}%
   {\csname oldend#1\endcsname\endlinenomath}%
}
\newcommand*\patchBothAmsMathEnvironmentsForLineno[1]{%
  \patchAmsMathEnvironmentForLineno{#1}%
  \patchAmsMathEnvironmentForLineno{#1*}%
}
\AtBeginDocument{%
\patchBothAmsMathEnvironmentsForLineno{equation}%
\patchBothAmsMathEnvironmentsForLineno{align}%
\patchBothAmsMathEnvironmentsForLineno{flalign}%
\patchBothAmsMathEnvironmentsForLineno{alignat}%
\patchBothAmsMathEnvironmentsForLineno{gather}%
\patchBothAmsMathEnvironmentsForLineno{multline}%
\patchBothAmsMathEnvironmentsForLineno{eqnarray}%
}

\usepackage{hyperxmp}

\usepackage[pdftex,
            pdfauthor={\paperauthors},
            pdftitle={\paperasciititle},
            pdfkeywords={\paperkeywords},
            pdfcopyright={Copyright (C) \papercopyright},
            pdflicenseurl={\paperlicenceurl}]{hyperref}

\usepackage[colorinlistoftodos,textsize=scriptsize]{todonotes}

\usepackage[bottom,flushmargin,hang,multiple]{footmisc}

\usepackage[all]{hypcap} 

\usepackage{xspace} 
\usepackage{upgreek}

\def\lhcb   {\mbox{LHCb}\xspace}

\def\MagUp {\mbox{\em Mag\kern -0.05em Up}\xspace}

\ifthenelse{\boolean{uprightparticles}}%
{

 \def\Pmu         {\ensuremath{\upmu}\xspace}

 \def\Ppi         {\ensuremath{\uppi}\xspace}                 
                  
 \def\Prho        {\ensuremath{\uprho}\xspace}

 \def\PDelta      {\ensuremath{\Delta}\xspace}                 
 \def\PXi         {\ensuremath{\Xi}\xspace}                 
 \def\PLambda     {\ensuremath{\Lambda}\xspace}                 
 \def\PSigma      {\ensuremath{\Sigma}\xspace}                 
 \def\POmega      {\ensuremath{\Omega}\xspace}                 
 \def\PUpsilon    {\ensuremath{\Upsilon}\xspace}
 \let\oldPi\Pi
 \def\PPi         {\ensuremath{\oldPi}\xspace}

 \def\PB      {\ensuremath{\mathrm{B}}\xspace}                 
                  
 \def\PD      {\ensuremath{\mathrm{D}}\xspace}

 \def\PK      {\ensuremath{\mathrm{K}}\xspace}

 \def\Pb      {\ensuremath{\mathrm{b}}\xspace}                 
 \def\Pc      {\ensuremath{\mathrm{c}}\xspace}                 
 \def\Pd      {\ensuremath{\mathrm{d}}\xspace}

 \def\Pi      {\ensuremath{\mathrm{i}}\xspace}

 \def\Pp      {\ensuremath{\mathrm{p}}\xspace}

 \def\Ps      {\ensuremath{\mathrm{s}}\xspace}                 
                  
 \def\Pu      {\ensuremath{\mathrm{u}}\xspace}

 \def\thebaroffset{0.0em}
}
{

 \def\Pmu         {\ensuremath{\mu}\xspace}

 \def\Ppi         {\ensuremath{\pi}\xspace}                 
                  
 \def\Prho        {\ensuremath{\rho}\xspace}

 \mathchardef\PDelta="7101
 \mathchardef\PXi="7104
 \mathchardef\PLambda="7103
 \mathchardef\PSigma="7106
 \mathchardef\POmega="710A
 \mathchardef\PUpsilon="7107
 \mathchardef\PPi="7105
                  
 \def\PB      {\ensuremath{B}\xspace}                 
                  
 \def\PD      {\ensuremath{D}\xspace}

 \def\PK      {\ensuremath{K}\xspace}

 \def\Pb      {\ensuremath{b}\xspace}                 
 \def\Pc      {\ensuremath{c}\xspace}                 
 \def\Pd      {\ensuremath{d}\xspace}

 \def\Pi      {\ensuremath{i}\xspace}

 \def\Pp      {\ensuremath{p}\xspace}

 \def\Ps      {\ensuremath{s}\xspace}                 
                  
 \def\Pu      {\ensuremath{u}\xspace}

 \def\thebaroffset{0.18em}
}
\newcommand{\offsetoverline}[2][\thebaroffset]{\kern #1\overline{\kern -#1 #2}}%

\makeatletter
\ifcase \@ptsize \relax
  \newcommand{\miniscule}{\@setfontsize\miniscule{4}{5}}
\or
  \newcommand{\miniscule}{\@setfontsize\miniscule{5}{6}}
\or
  \newcommand{\miniscule}{\@setfontsize\miniscule{5}{6}}
\fi
\makeatother

\DeclareRobustCommand{\optbar}[1]{\shortstack{{\miniscule (\rule[.5ex]{1.25em}{.18mm})}
  \\ [-.7ex] $#1$}}

\def\mup        {{\ensuremath{\Pmu^+}}\xspace}
\def\mun        {{\ensuremath{\Pmu^-}}\xspace} 

\def\uquark    {{\ensuremath{\Pu}}\xspace}

\def\dquark    {{\ensuremath{\Pd}}\xspace}

\def\squark    {{\ensuremath{\Ps}}\xspace}

\def\cquark    {{\ensuremath{\Pc}}\xspace}

\def\bquark    {{\ensuremath{\Pb}}\xspace}

\def\pion   {{\ensuremath{\Ppi}}\xspace}
\def\piz    {{\ensuremath{\pion^0}}\xspace}
\def\pip    {{\ensuremath{\pion^+}}\xspace}
\def\pim    {{\ensuremath{\pion^-}}\xspace}

\def\rhomeson {{\ensuremath{\Prho}}\xspace}

\def\rhop     {{\ensuremath{\rhomeson^+}}\xspace}

\def\kaon    {{\ensuremath{\PK}}\xspace}

\def\KorKbar {\kern \thebaroffset\optbar{\kern -\thebaroffset \PK}{}\xspace}
\def\Kz      {{\ensuremath{\kaon^0}}\xspace}

\def\Kp      {{\ensuremath{\kaon^+}}\xspace}
\def\Km      {{\ensuremath{\kaon^-}}\xspace}
\def\Kpm     {{\ensuremath{\kaon^\pm}}\xspace}

\def\Kstarz  {{\ensuremath{\kaon^{*0}}}\xspace}

\def\Kstar   {{\ensuremath{\kaon^*}}\xspace}

\def\Kstarp  {{\ensuremath{\kaon^{*+}}}\xspace}

\def\Kstarpm {{\ensuremath{\kaon^{*\pm}}}\xspace}

\def\Dbar    {{\ensuremath{\offsetoverline{\PD}}}\xspace}
\def\D       {{\ensuremath{\PD}}\xspace}

\def\DorDbar {\kern \thebaroffset\optbar{\kern -\thebaroffset \PD}\xspace}
\def\Dz      {{\ensuremath{\D^0}}\xspace}
\def\Dzb     {{\ensuremath{\Dbar{}^0}}\xspace}
\def\Dp      {{\ensuremath{\D^+}}\xspace}
\def\Dm      {{\ensuremath{\D^-}}\xspace}

\def\DpDm    {\ensuremath{\Dp {\kern -0.16em \Dm}}\xspace}
\def\Dstar   {{\ensuremath{\D^*}}\xspace}

\def\Dstarz  {{\ensuremath{\D^{*0}}}\xspace}
\def\Dstarzb {{\ensuremath{\Dbar{}^{*0}}}\xspace}
\def\theDstarz{{\ensuremath{\D^{*}(2007)^{0}}}\xspace}
\def\theDstarzb{{\ensuremath{\Dbar^{*}(2007)^{0}}}\xspace}
\def\Dstarp  {{\ensuremath{\D^{*+}}}\xspace}
\def\Dstarm  {{\ensuremath{\D^{*-}}}\xspace}

\def\Ds      {{\ensuremath{\D^+_\squark}}\xspace}

\def\Dsm     {{\ensuremath{\D^-_\squark}}\xspace}

\def\Dsmp    {{\ensuremath{\D^{\mp}_\squark}}\xspace}

\def\B       {{\ensuremath{\PB}}\xspace}
\def\Bbar    {{\ensuremath{\offsetoverline{\PB}}}\xspace}

\def\BorBbar {\kern \thebaroffset\optbar{\kern -\thebaroffset \PB}\xspace}
\def\Bz      {{\ensuremath{\B^0}}\xspace}

\def\Bd      {{\ensuremath{\B^0}}\xspace}

\def\BdorBdbar {\kern \thebaroffset\optbar{\kern -\thebaroffset \Bd}\xspace}

\def\Bub     {{\ensuremath{\B^-}}\xspace}

\def\Bm      {{\ensuremath{\Bub}}\xspace}

\def\Bs      {{\ensuremath{\B^0_\squark}}\xspace}
\def\Bsb     {{\ensuremath{\Bbar{}^0_\squark}}\xspace}
\def\BsorBsbar {\kern \thebaroffset\optbar{\kern -\thebaroffset \Bs}\xspace}

\def\Bds     {{\ensuremath{\B_{(\squark)}^0}}\xspace}

\def\Y#1S{\ensuremath{\PUpsilon{(#1S)}}\xspace}

\def\proton      {{\ensuremath{\Pp}}\xspace}
\def\antiproton  {{\ensuremath{\overline \proton}}\xspace}

\def\Lz          {{\ensuremath{\PLambda}}\xspace}
\def\Lbar        {{\ensuremath{\offsetoverline{\PLambda}}}\xspace}
\def\LorLbar     {\kern \thebaroffset\optbar{\kern -\thebaroffset \PLambda}\xspace}

\def\Xires       {{\ensuremath{\PXi}}\xspace}

\def\Xiresbar       {{\ensuremath{\offsetoverline{\Xires}}}\xspace}

\def\Lb           {{\ensuremath{\Lz^0_\bquark}}\xspace}
\def\Lbbar        {{\ensuremath{\Lbar{}^0_\bquark}}\xspace}

\def\Xibbar       {{\ensuremath{\Xiresbar{}_\bquark}}\xspace}

\def\BF         {{\ensuremath{\mathcal{B}}}\xspace}
\def\BR         {\BF}

\newcommand{\decay}[2]{\ensuremath{#1\!\to #2}\xspace} 

\def\to                 {\ensuremath{\rightarrow}\xspace}

\def\CP                {{\ensuremath{C\!P}}\xspace}

\def\Vud  {{\ensuremath{V_{\uquark\dquark}^{\phantom{\ast}}}}\xspace}
\def\Vcd  {{\ensuremath{V_{\cquark\dquark}^{\phantom{\ast}}}}\xspace}

\def\Vubs  {{\ensuremath{V_{\uquark\bquark}^\ast}}\xspace}
\def\Vcbs  {{\ensuremath{V_{\cquark\bquark}^\ast}}\xspace}

\def\AT#1     {\ensuremath{A_{\mathrm{T}}^{#1}}\xspace}           

\def\C#1      {\ensuremath{\mathcal{C}_{#1}}\xspace}                       
\def\Cp#1     {\ensuremath{\mathcal{C}_{#1}^{'}}\xspace}                    
\def\Ceff#1   {\ensuremath{\mathcal{C}_{#1}^{\mathrm{(eff)}}}\xspace}        
\def\Cpeff#1  {\ensuremath{\mathcal{C}_{#1}^{'\mathrm{(eff)}}}\xspace}       
\def\Ope#1    {\ensuremath{\mathcal{O}_{#1}}\xspace}                       
\def\Opep#1   {\ensuremath{\mathcal{O}_{#1}^{'}}\xspace}                    

\newcommand{\nospaceunit}[1]{\ensuremath{\text{#1}}}       
\newcommand{\aunit}[1]{\ensuremath{\text{\,#1}}}       

\newcommand{\tev}{\aunit{Te\kern -0.1em V}\xspace}
\newcommand{\gev}{\aunit{Ge\kern -0.1em V}\xspace}
\newcommand{\mev}{\aunit{Me\kern -0.1em V}\xspace}
\newcommand{\kev}{\aunit{ke\kern -0.1em V}\xspace}
\newcommand{\ev}{\aunit{e\kern -0.1em V}\xspace}
 
\newcommand{\mevc}{\ensuremath{\aunit{Me\kern -0.1em V\!/}c}\xspace}
\newcommand{\gevc}{\ensuremath{\aunit{Ge\kern -0.1em V\!/}c}\xspace}
\newcommand{\mevcc}{\ensuremath{\aunit{Me\kern -0.1em V\!/}c^2}\xspace}
\newcommand{\gevcc}{\ensuremath{\aunit{Ge\kern -0.1em V\!/}c^2}\xspace}

\def\mum  {\ensuremath{\,\upmu\nospaceunit{m}}\xspace}

\def\fb   {\ensuremath{\aunit{fb}}\xspace}
\def\invfb   {\ensuremath{\fb^{-1}}\xspace}

\newcommand{\chisq}{\ensuremath{\chi^2}\xspace}
\newcommand{\chisqndf}{\ensuremath{\chi^2/\mathrm{ndf}}\xspace}
\newcommand{\chisqip}{\ensuremath{\chi^2_{\text{IP}}}\xspace}

\def\deriv {\ensuremath{\mathrm{d}}}

\def\gsim{{~\raise.15em\hbox{$>$}\kern-.85em
          \lower.35em\hbox{$\sim$}~}\xspace}
\def\lsim{{~\raise.15em\hbox{$<$}\kern-.85em
          \lower.35em\hbox{$\sim$}~}\xspace}

\def\pt         {\ensuremath{p_{\mathrm{T}}}\xspace}

\def\ptot       {\ensuremath{p}\xspace}

\def\rad{\aunit{rad}\xspace}

\def\evtgen     {\mbox{\textsc{EvtGen}}\xspace}

\def\geant      {\mbox{\textsc{Geant4}}\xspace}

\def\photos     {\mbox{\textsc{Photos}}\xspace}

\def\pythia     {\mbox{\textsc{Pythia}}\xspace}

\def\tell1  {TELL1\xspace}
\def\ukl1   {UKL1\xspace}

\newcommand{\eg}{\mbox{\itshape e.g.}\xspace}

\newcommand{\lhcborcid}[1]{\href{https://orcid.org/#1}{\hspace*{0.1em}\raisebox{-0.45ex}{\includegraphics[width=1em]{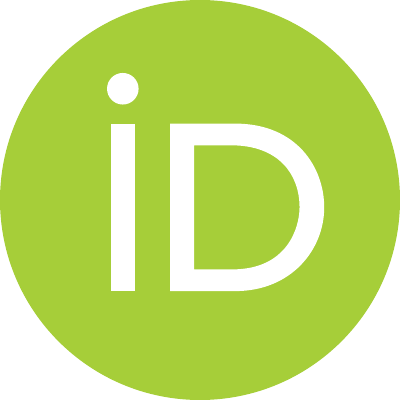}}}}

\usepackage{cite} 
\usepackage{mciteplus}

\usepackage{longtable} 

\begin{document}

\renewcommand{\thefootnote}{\fnsymbol{footnote}}
\setcounter{footnote}{1}


\begin{titlepage}
\pagenumbering{roman}

\vspace*{-1.5cm}
\centerline{\large EUROPEAN ORGANIZATION FOR NUCLEAR RESEARCH (CERN)}
\vspace*{1.5cm}
\noindent
\begin{tabular*}{\linewidth}{lc@{\extracolsep{\fill}}r@{\extracolsep{0pt}}}
\ifthenelse{\boolean{pdflatex}}
{\vspace*{-1.5cm}\mbox{\!\!\!\includegraphics[width=.14\textwidth]{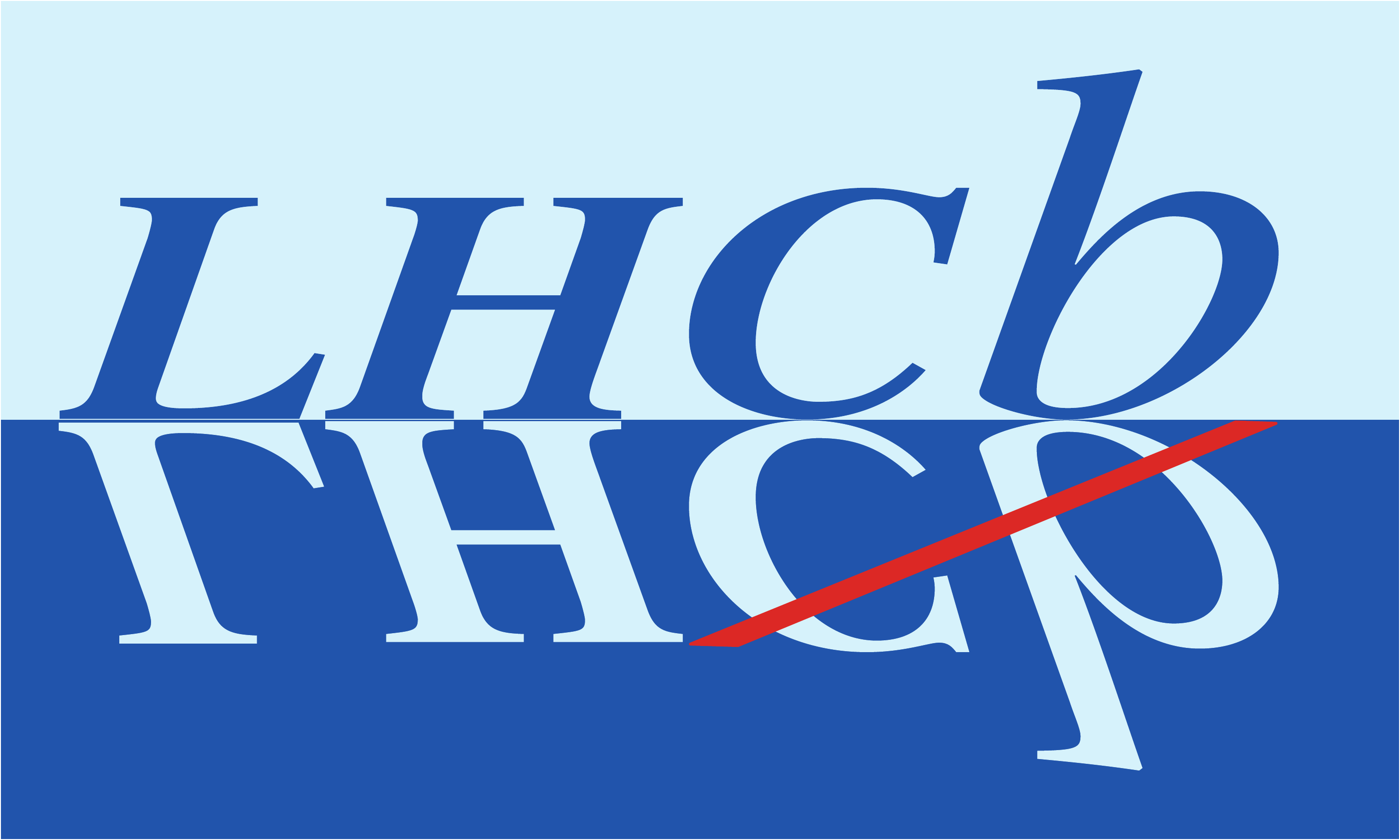}} & &}%
{\vspace*{-1.2cm}\mbox{\!\!\!\includegraphics[width=.12\textwidth]{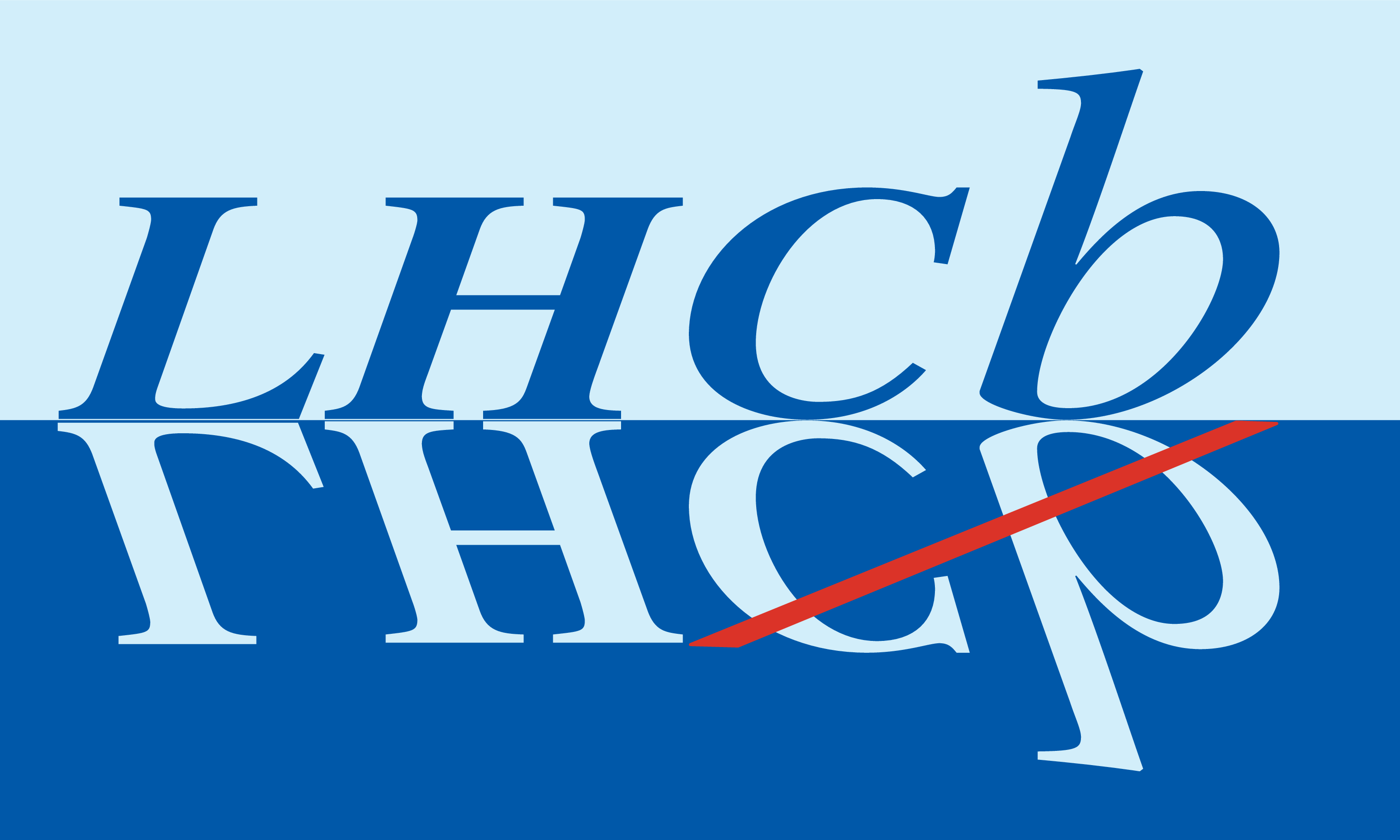}} & &}%
\\
 & & CERN-EP-2023-140 \\  
 & & LHCb-PAPER-2023-014 \\  
 & & August 1, 2023 \\ 
\end{tabular*}

\vspace*{1.5cm}

{\normalfont\bfseries\boldmath\huge
\begin{center}
  \papertitle 
\end{center}
}

\vspace*{0.5cm}

\begin{center}
\paperauthors\footnote{Authors are listed at the end of this paper.}
\end{center}


\begin{abstract}
  \noindent
  This paper reports the observation of the decays \BdsDsoK 
  using proton-proton collision data collected by the \lhcb experiment, corresponding to an integrated luminosity of $9\invfb$.
  The branching fractions of these decays are measured relative to the normalisation channel \BdDKK. 
  The \Dsom meson is reconstructed in the $\theDstarzb\Km$ decay channel and the products of branching fractions are measured to be
  \begin{align*}
    \centering
    \BF(\decay{\Bs}{\Dsomp\Kpm})&\times \BF(\decay{\Dsom}{\theDstarzb\Km}) \\
    &= \ResultBRBsDsoKDstK,\\
    \BF(\decay{\Bz}{\Dsomp\Kpm})&\times \BF(\decay{\Dsom}{\theDstarzb\Km}) \\
    &= \ResultBRBzDsoKDstK.
  \end{align*}
  The first uncertainty is statistical, the second systematic, and the third arises from the uncertainty of the branching fraction of the \decay{\Bz}{\Dzb\Kp\Km} normalisation channel.
  The last uncertainty in the \Bs result is due to the limited knowledge of the fragmentation fraction ratio, $f_s/f_d$. The significance for the \Bs and \Bd signals is larger than $10\,\sigma$.
  The ratio of the helicity amplitudes which governs the angular distribution of the $\decay{\Dsom}{\theDstarzb\Km}$ decay is determined from the data. The ratio of the $S$- and $D$-wave amplitudes is found to be $\ResultA$ and the phase difference between them $\ResultB$ rad, where the first uncertainty is statistical and the second systematic.
\end{abstract}

\vspace*{1.0cm}

\begin{center}
  Published in
  JHEP 10 (2023) 106
\end{center}

\vspace{\fill}

{\footnotesize 
\centerline{\copyright~\papercopyright. \href{\paperlicenceurl}{\paperlicence}.}}
\vspace*{2mm}

\end{titlepage}


\newpage
\setcounter{page}{2}
\mbox{~}
%
%
%
%


\renewcommand{\thefootnote}{\arabic{footnote}}
\setcounter{footnote}{0}

\cleardoublepage


\pagestyle{plain} 
\setcounter{page}{1}
\pagenumbering{arabic}


\section{Introduction}
\label{sec:Introduction}
\begin{figure}[!b]
    \begin{center}
        \includegraphics[width=0.45\textwidth]{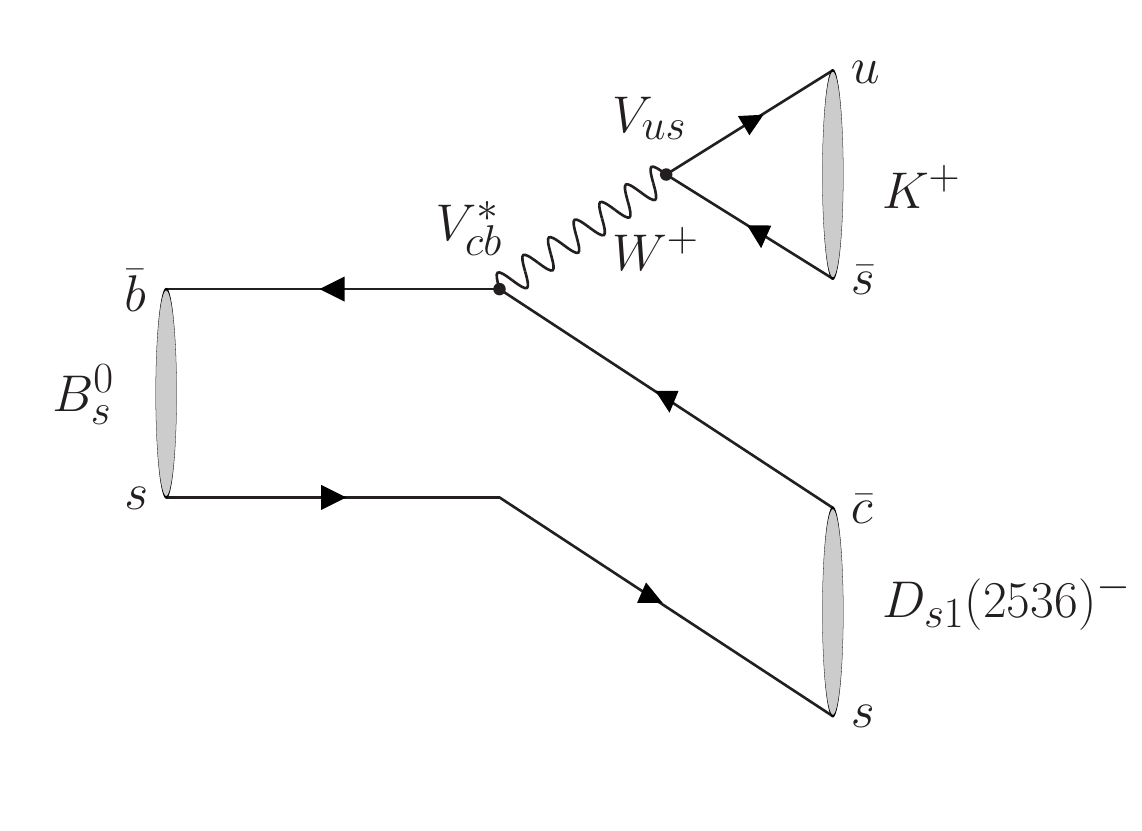}
        \includegraphics[width=0.45\textwidth]{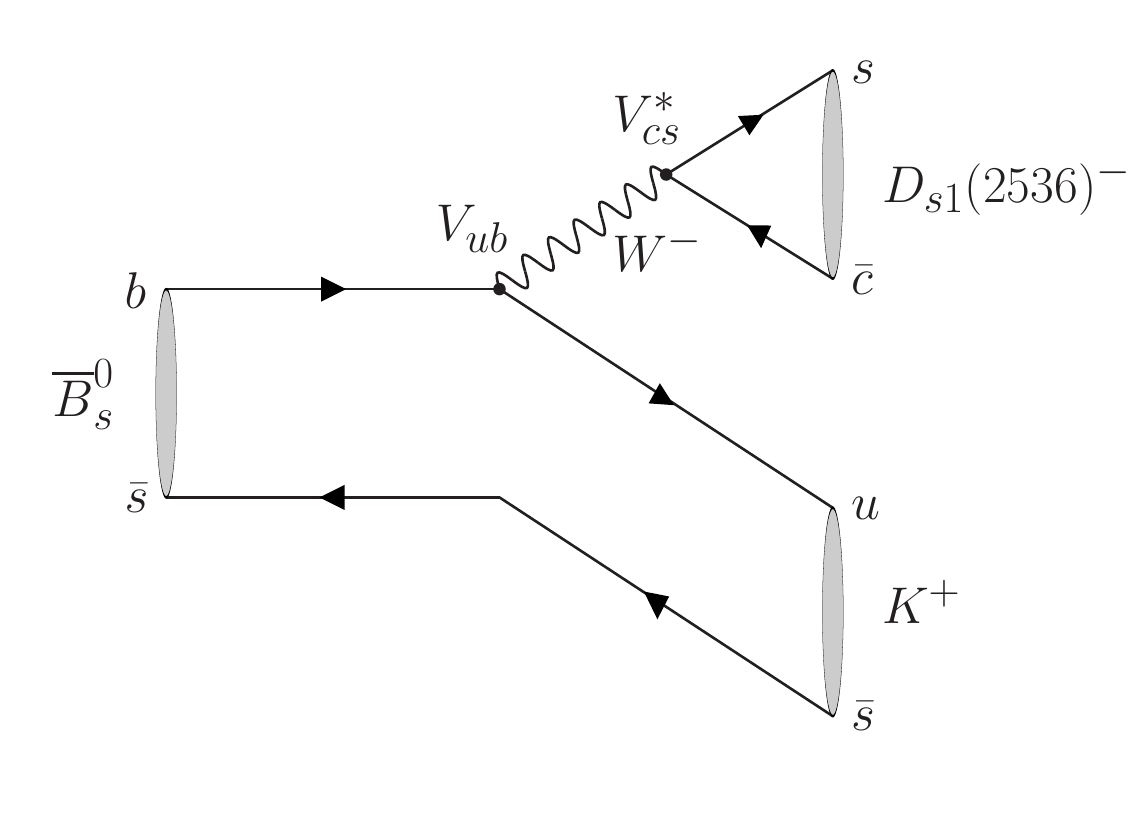}
    \end{center}
    \vspace{-0.5cm}
    \caption{Feynman diagrams for the \decay{\Bs}{\Dsom\Kp} (left)  and \decay{\Bsb}{\Dsom\Kp} (right) decay channel.}
    \label{fig:intro:Feynman_diagram}
\end{figure}

Precise measurements of \CP violation are essential tests of the Standard Model (SM) of particle physics.
In the SM, \CP violation originates from a single phase in the  Cabibbo-Kobayashi-Maskawa (CKM) matrix\cite{Cabibbo:1963yz, Kobayashi:1973fv}.
The angle $\gamma \equiv \text{arg}(-\Vud\Vubs/\Vcd\Vcbs)$, one of the parameters related to this \CP violation phase, can be measured at the tree level using the interference between $\decay{\bquark}{\cquark\squark\bar{\uquark}}$ and $\decay{\bquark}{\uquark\squark\bar{\cquark}}$ transitions. 
The comparison of this measurement with the value determined using other measurements involving particle loops serves as a probe to search for physics beyond the SM. 
Previous \lhcb analyses of \decay{\Bs}{\Dsmp\Kpm}~\cite{LHCb-PAPER-2017-047} and \decay{\Bs}{\Dsm\Kp\pip\pim}~\cite{LHCB-PAPER-2020-030} decays have reported the value of $\gamma$ based on \Bs decays to final states with a \Ds meson and one or more light mesons. The unobserved \Bs decay modes, $\BsDsoK$, have a similar topology. Their Feynman diagrams are shown in Fig.~\ref{fig:intro:Feynman_diagram}. 
The $\BsDsoK$ decay adds a further channel through which $\gamma$ can be determined using a time-dependent method.

Additionally, studies of the \BdsDsoK decays are also helpful to understand the pattern in the decay modes \decay{\Bz}{\D^{(*)-}\Kp} and \decay{\Bs}{\D_s^{(*)-}\pip},
 whose measured branching fractions~\cite{PDG2022} are smaller than the predictions with QCD factorisation~\cite{Fleischer:2010ca,Bordone:2020gao,
Cai:2021mlt,Endo:2021ifc}. 
In Ref.~\cite{Fleischer:2021cct} the branching fractions of the \decay{\Bs}{\Dsmp\Kpm} decays are extracted by combining the \CP-violating observables with the measured average branching fraction of these two decays, taking into account the effects of $\Bs$-$\Bsb$ mixing.
The result shows a similar difference with the calculations based on QCD factorisation. The \BdsDsoK decays can provide additional information to validate this difference between experimental and theoretical results thanks to the similar topology as the \decay{\Bs}{\Dsmp\Kpm} decays.

This paper reports the observation of the decays \BdsDsoK and a measurement of their branching fractions using proton-proton~($pp$) collision data collected with the \lhcb detector.
The data were collected at centre-of-mass energies of $7\tev$, $8\tev$ and $13\tev$, corresponding to integrated luminosities of $1\invfb$, $2\invfb$ (Run 1), and $6\invfb$ (Run 2), respectively. In \BdsDsoK decays, the \Dsom meson is reconstructed in its decay to the \theDstarzb\Km state, with the \theDstarzb meson  partially reconstructed by both $\Dzb\gamma$ and $\Dzb\piz$ decays, where the low-momentum neutral particles, $\gamma$ and $\piz$, are not reconstructed. The \Dzb meson is reconstructed by the Cabibbo-favoured (CF) decay mode $\Dzb\to\Kp\pim$. Throughout this paper, charge conjugation is implied unless explicitly stated otherwise, and the symbol \Dstarz is used to denote the $\theDstarz$ meson.

To measure the branching fractions of ${\BdsDsoK}$ decays, the ${\BdDKK}$ decay is used as a normalisation channel. The ratio of branching fractions for the decays \BdsDsoK with the secondary decay \decay{\Dsom}{\Dstarzb\Km} to \BdDKK is defined as 
\begin{equation}
\label{eq:01:01}
\calR(X) \equiv \frac{\BR(X)\times \BR(\decay{\Dsom}{\Dstarzb\Km})}{\BR(\decay{\Bz}{\Dzb\Kp\Km})},
\end{equation}
where $X$ indicates the \mbox{\decay{\Bds}{\Dsomp\Kpm}} decay channels.
For the ${\decay{\Dsom}{\Dstarzb\Km}}$ decay, the ratio of its helicity amplitudes is measured and based on that, the ratio of the $S$- and $D$-wave amplitudes is also obtained.

\section{Angular decay rate formalism}
\label{sec:angular_decay_rate}
The decays \BdsDsomK involve the production of one pseudovector, $\Dsom$, and one pseudoscalar meson, $\Kp$, from a pseudoscalar \Bds parent. 
The subsequent process, $\decay{\Dsom}{\Dstarzb\Km}$, presents a polarisation structure where three complex helicity amplitudes,  $H_0$, $H_+$ and $H_-$, contribute to the total decay rate. 
These three amplitudes correspond to the orientation of the linear polarisation of the vector particle, \Dstarzb, with respect to the pseudovector meson, \Dsom. The two transverse amplitudes, $H_+$ and $H_-$,  describe the helicity value of $1$ and $-1$ of the \Dstarzb meson and the longitudinal amplitude, $H_0$, the null helicity value. Parity conservation implies $H_+$ equal to $H_-$.

\begin{figure}[!t]
    \centering
    \includegraphics[width=0.25\textwidth]{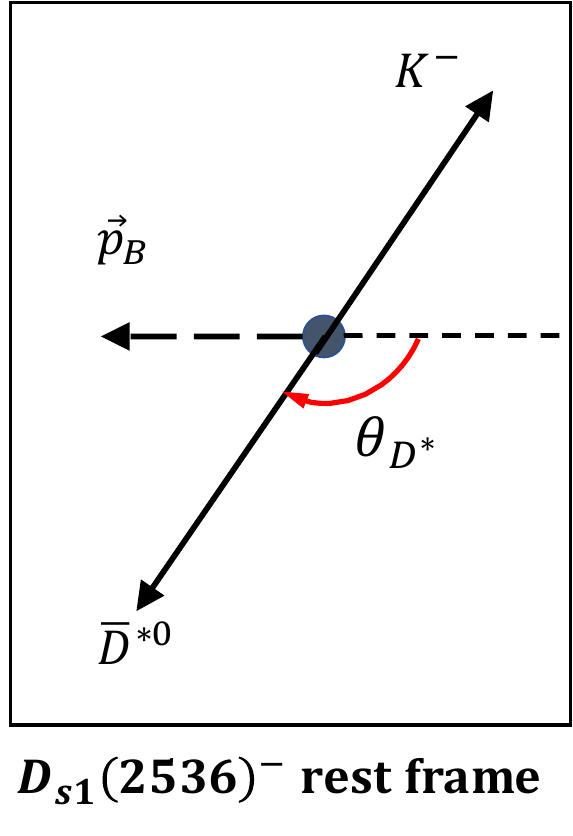}
    \includegraphics[width=0.246\textwidth]{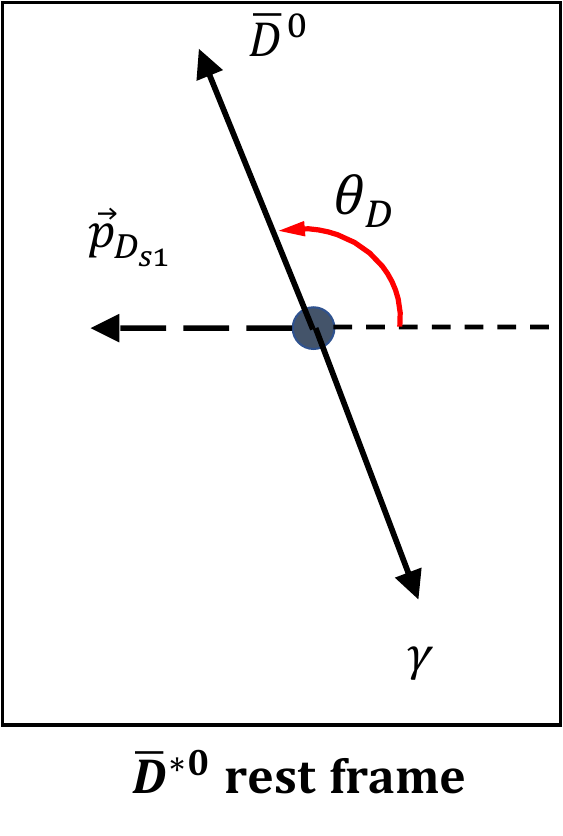}
    \includegraphics[width=0.4\textwidth]{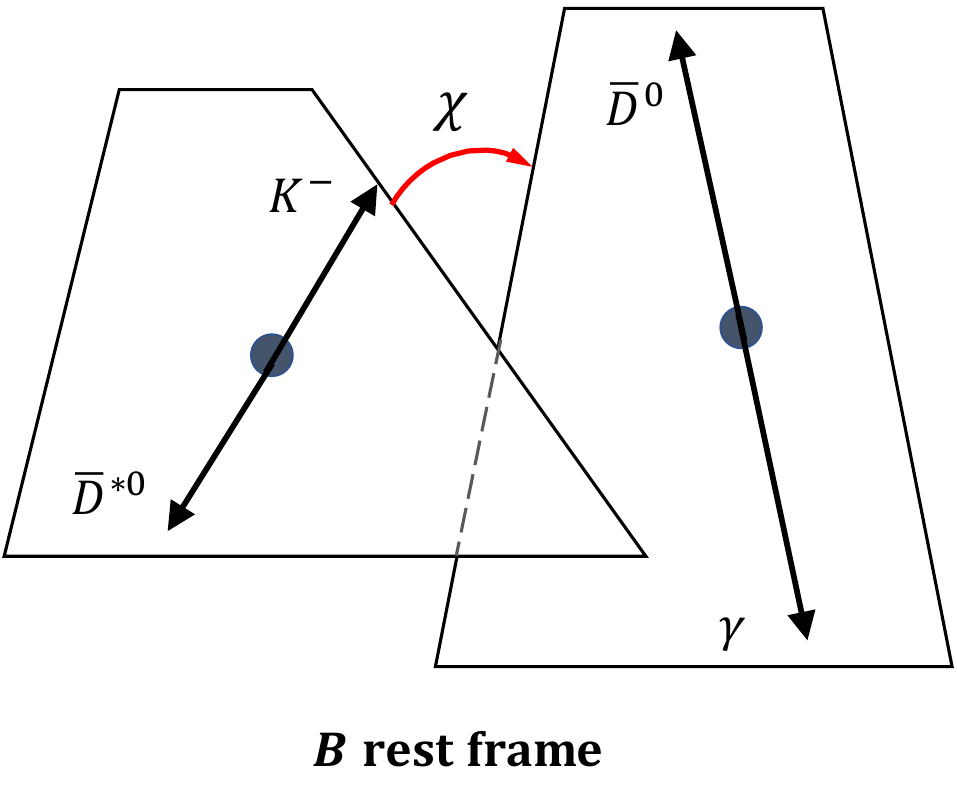}
    \caption{Illustration of the \BdsDsoK decay angles.}
    \label{fig:definition_of_angles}
\end{figure}

The decay rates of \BdsDsoK are a function of three decay angles, $\theta_{\Dstar}$, $\theta_{\D}$ and $\chi$ (see Fig.~\ref{fig:definition_of_angles}): $\theta_{\Dstar}$ is the angle between the \Dstarzb meson and the direction opposite to the \B momentum vector in the \Dsom rest frame, $\theta_{\D}$ is the angle between the \Dzb meson and the direction opposite to the \Dso momentum vector in the \Dstarzb rest frame, and $\chi$ is the angle between the two decay planes defined in the \B rest frame. 
The full three-dimensional differential decay rate expressed in terms of the helicity amplitudes is 
\begin{equation}
    \label{equ:angular_decay_rate:differential_decay_rate}
    \begin{split}
        \frac{\deriv^3 \Gamma}{\deriv \cos \theta_{\Dstar} \deriv \cos \theta_{\D} \deriv \chi} &\propto \omega_{\textrm{long}}(\theta_\Dstar, \theta_\D) |H_0|^2\\ &+ \omega_\textrm{tran}(\chi, \theta_\Dstar, \theta_\D) |H_+|^2 + \omega_\textrm{int}(\chi, \theta_\Dstar, \theta_\D) \Re(H_0^* H_+),\\
    \end{split}
\end{equation}
where the factors $\omega_\textrm{long}$, $\omega_\textrm{tran}$ and $\omega_\textrm{int}$ are functions of the decay angles associated with the longitudinal component, the transverse component and the interference term between the longitudinal and transverse components, respectively. The ratio of the amplitudes of longitudinal and transverse components, $H_+/H_0$, is expressed as  $ke^{i\phi}$ where $k>0$ and $\phi\in[-\pi, \pi]$. 
The interference term, $\Re(H_0H_+)$, is only sensitive to $\cos(\phi)$, therefore $\phi$ has a two-fold ambiguity, and only the absolute value $|\phi|$ is reported. 
In addition, the signal channel is split into the photon chain (with the intermediate decay \decay{\Dstarzb}{\Dzb\gamma}) and the \piz chain (with the decay \decay{\Dstarzb}{\Dzb\piz}). 
The $\omega$ factors and helicity amplitudes are different for the photon and \piz chains, but they share the same $k$ and $\phi$. 
The expressions for the $\omega$ factors for the two chains are given in Table~\ref{tab:angular_decay_rate:omega_factor}. 
This formalism is the same as that adopted in other \lhcb angular analyses such as that of \decay{\B}{\Kstar\mup\mun} decays \cite{LHCb-PAPER-2015-051,LHCb-PAPER-2020-041}.
For the decay chain, $\decay{\Dsom}{\Dstarzb\Km}, \decay{\Dstarzb}{\Dzb(\piz/\gamma)}$, the decay angle $\theta_{\Dstar}$ and the orientation of the decay plane of $\Dstarzb\Km$ are uniform in their value range. 
Therefore, the differential rate of the $\Dsom$ decay is obtained by integrating Eq.~\ref{equ:angular_decay_rate:differential_decay_rate} over $\theta_{\Dstar}$ and $\chi$, resulting in no contribution from $\omega_{\textrm{int}}$. However, for \BdsDsoK decays, there is \textcolor{blue}{a} non-zero interference contribution in the decay rate.

\begin{table}[!b]
    \centering
    \caption{Definitions of the $\omega$ functions in the differential decay rate.}
    \label{tab:angular_decay_rate:omega_factor}
    \begin{tabular}{l|c|c}
    \toprule[1pt]
         & photon chain & $\piz$ chain\\
         \hline
         $\omega_{\textrm{long}}(\theta_\Dstar, \theta_\D)$&  $\cos^2\theta_\Dstar \sin^2\theta_\D$ & $\cos^2\theta_\Dstar \cos^2\theta_\D$\\  
         $\omega_\textrm{tran}(\chi, \theta_\Dstar, \theta_\D)$ & $\sin^2\theta_\Dstar (\sin^2\chi + \cos^2\chi \cos^2\theta_D)$ & $\cos^2\chi \sin^2\theta_\Dstar \sin^2\theta_\D$ \\
         $\omega_\textrm{int}(\chi, \theta_\Dstar, \theta_\D)$ & $2\cos\chi \sin\theta_\Dstar \cos\theta_\Dstar \sin\theta_\D \cos\theta_\D$ & $-2\cos\chi \sin\theta_\Dstar \cos\theta_\Dstar \sin\theta_\D \cos\theta_\D$\\
         \bottomrule[1pt]
    \end{tabular}
\end{table}

\section{\lhcb detector}
\label{sec:LHCb_detector}
The \lhcb detector~\cite{LHCb-DP-2008-001,LHCb-DP-2014-002} is a single-arm forward
spectrometer covering the \mbox{pseudorapidity} range $2<\eta <5$,
designed for the study of particles containing \bquark or \cquark
quarks. The detector includes a high-precision tracking system
consisting of a silicon-strip vertex detector surrounding the $pp$
interaction region, a large-area silicon-strip detector located
upstream of a dipole magnet with a bending power of about
$4{\mathrm{\,Tm}}$, and three stations of silicon-strip detectors and straw
drift tubes placed downstream of the magnet.
The tracking system provides a measurement of the momentum, \ptot, of charged particles with
a relative uncertainty that varies from 0.5\% at low momentum to 1.0\% at 200\gevc.
The minimum distance of a track to a primary $pp$ collision vertex (PV), the impact parameter (IP), is measured with a resolution of $(15+29/\pt)\mum$, where \pt is the component of the momentum transverse to the beam, in\,\gevc.
Different types of charged hadrons are distinguished using information
from two ring-imaging Cherenkov detectors. 
Photons, electrons and hadrons are identified by a calorimeter system consisting of
scintillating-pad and preshower detectors, an electromagnetic calorimeter and a hadronic calorimeter. Muons are identified by a system composed of alternating layers of iron and multiwire proportional chambers.

The online event selection is performed by a trigger, which consists of a hardware stage, based on information from the calorimeter and muon systems, followed by a software stage, which applies a full event reconstruction. 
At the hardware trigger stage, events are required to have a muon with high \pt or a hadron, photon or electron with high transverse energy in the calorimeters. For hadrons, the transverse energy threshold is $3.5\gev$.
A global hardware trigger decision is ascribed to the reconstructed candidate, the rest of the event or a combination of both; events triggered as such are defined respectively as triggered on signal (TOS), triggered independently of signal (TIS), and triggered on both.
The software trigger requires a two-, three- or four-track secondary vertex with a significant displacement from any primary $pp$ interaction vertex. At least one charged particle must have a transverse momentum $\pt > 1.6\gevc$ and be inconsistent with originating from a PV. A multivariate algorithm~\cite{BBDT,LHCb-PROC-2015-018} is used for the identification of secondary vertices consistent with the decay of a \bquark hadron.

In the simulation, $pp$ collisions are generated using \pythia~\cite{Sjostrand:2007gs,*Sjostrand:2006za} with a specific \lhcb configuration~\cite{LHCb-PROC-2010-056}. Decays of unstable particles are described by \evtgen~\cite{Lange:2001uf}, in which final-state radiation is generated using \photos~\cite{davidson2015photos}. The interaction of the generated particles with the detector, and its response, are implemented using the \geant toolkit~\cite{Allison:2006ve, *Agostinelli:2002hh} as described in Ref.~\cite{LHCb-PROC-2011-006}.

\section{Event selection}
\label{sec:event_selection}

For the signal channels, the \Bds candidate is reconstructed from a \Dsom candidate and an additional kaon of opposite charge. 
The  \Dsom candidate is reconstructed through the \Dstarzb\Km mode, and the \Dstarzb candidate is partially reconstructed in the $\Dzb\gamma$ or $\Dzb\piz$ modes, where the photon or $\piz$ is not reconstructed. 
For the normalisation channel, the \Bd candidate is reconstructed from a \Dzb candidate and two kaon candidates of opposite charge. 
For both signal and normalisation channels, the \Dzb candidate is reconstructed in the Cabbibo favoured decay, \decay{\Dzb}{\Kp\pim}.
The selection criteria discussed below are applied to both signal and normalisation channels, except if explicitly mentioned otherwise.

The tracks of the final-state particles are required to be of good fit quality and have a high $\chisqip$ value with respect to any PV, where $\chisqip$ is defined as the difference in the vertex-fit $\chisq$ of a given PV reconstructed with and without the particle being considered. Tracks are also required to be within the kinematic coverage of the RICH detectors, which provide particle identification (PID) information used to reduce backgrounds with mis-identified particles. The kaon track from the \Bds vertex, which has the same charge as the kaon from the \Dzb candidate, is required to fulfil stringent PID requirements, and the other three final-state particles are selected with loose PID requirements. The reconstructed mass of the \Dzb meson is required to be within $25 \mevcc$ of its known value \cite{PDG2022}, which reduces background from decays with mis-identified hadrons such as \decay{\Dzb}{\Kp\Km(\pip\pim)}, in combination with the PID requirements on the \Dzb decay particles. The \Dzb decay vertex is required to be well reconstructed, with $\chisqndf<6$, where $\text{ndf}$ is the number of degrees of freedom. A vertex formed by one \Dzb and two kaon candidates with $\chisqndf<4$ is required to originate from a \Bds candidate. The reconstructed \Dzb and \Bds vertices are required to be significantly displaced from the associated PV, defined as the PV with the smallest $\chisqip$ with respect to the \Bds candidate, in case of more than one PV in the $pp$ collision.
In order to improve the \Bds mass resolution, a kinematic fit~\cite{Hulsbergen:2005pu} is performed with the \Dzb mass constrained to its known mass~\cite{PDG2022} and the \Bds momentum constrained to point to the PV with the smallest $\chisqip$. The \Bds candidates with invariant mass in the interval $5000-6000\mevcc$ are retained.

A multilayer perceptron (MLP) neural network technique, provided by the TMVA package \cite{Hocker:2007ht}, is used to further reduce the combinatorial background, consisting of random combination of tracks that mimic the signal, while preserving a high signal efficiency. 
The MLP classifier is trained with a simulated sample of the signal, and data from the upper $m(\Dzb\Kp\Km)$ sideband to represent combinatorial background.
The variables used in the training procedure are
the smallest \chisqip with respect to the PV and \pt of the decay products from the \B-decay vertex;
the \chisq probability of the $B$ vertex fit; 
the \chisq probability of the distance from the PV to the \B-decay vertex;
and the signed minimum cosine of the angle between the direction of one of the charged tracks from the $B$ decay and the $\Dzb$ meson, as projected in the plane perpendicular to the beam axis.
The MLP neural network is trained separately for Run 1 and Run 2 data samples and applied to both the \BdsDsoK and \BdDKK decays. The threshold for the MLP response is chosen by optimising the significance of the \Bs meson signal yield with the help of simulated samples, which reduces the combinatorial background by an order of magnitude while retaining about 88\% of the signal.

Considering that the invariant mass of the \Dsom meson is $2535\mevcc$ \cite{PDG2022} and it is partially reconstructed, the invariant mass of the \Dzb\Km pair is far below $2500\mevcc$. Thus, a requirement of $m(\Dzb\Km) < 2500 \mevcc$ is  applied to the signal channels, which keeps almost all signal candidates and reduces the background by about 95\%.
Other requirements are additionally applied to the \BdDKK normalisation channel.
The significance of the distance in the $z$ direction along the beam axis, defined as $(z_{\D}-z_{\B})/\sqrt{\sigma_{z_\D}^2 + \sigma_{z_\B}^2}$, where $z_{\D(\B)}$ and $\sigma_{z_{\D(\B)}}$ are the $z$ position and its uncertainty along the $z$ axis of the decay vertex of $\Dz(\Bds)$, is used to remove the background from charmless \Bds decays, such as \decay{\Bds}{\Kp\pim h^+ h^-} decays ($h$ stands for any light hadron). 
To remove the large background from \decay{\Bz}{\Dstarm\pip} decays, the mass difference $m(\Dzb\pim)-m(\Dzb)$ must not lie within $2.5\mevcc$ of the known $\Dstarm-\Dzb$ mass difference~\cite{PDG2022}. 
Moreover, the swapped background, where the $\Kpm$ from the \Dzb meson decay and the $\Kpm$ of the \Bds meson decay are swapped during reconstruction, is vetoed if its reconstructed mass with the two kaons exchanged lies in the range within $25\mevcc$ around the known \Dzb mass.

If there is more than one candidate existing in one $pp$ collision, only the candidate with the smallest value of the sum of \chisqip for the \Bds and \Dzb vertices, or the largest PID accuracy if the sum of \chisqip is the same, is retained; the associated systematic uncertainty for this choice is found to be negligible.

\section{Mass fit}
\label{sec:mass_fit}

The $m(\Dzb\Km)$ and $m(\Dzb\Km\Kp)$ distributions are used to distinguish signal from background. As neutral particles, photon or \piz, are not reconstructed, \BdsDsoK decays cannot be described by Gaussian functions in these two mass distributions, and are decomposed into three components: longitudinal~($H_+$), transverse~($H_0$) and interference. 
The shapes of $m(\Dzb\Km\Kp)$ distribution for longitudinal, transverse and interference components in each chain are obtained by weighting the simulated sample, uniformly generated in the phase space, with the corresponding $\omega$ factors defined in Table~\ref{tab:angular_decay_rate:omega_factor}.
The resulting shapes, $S_{X,\text{long}}, S_{X,\text{tran}}$ and $S_{X,\text{int}}$, for decay channel $X$, can be written as
\begin{equation}
    \begin{split}
        \sum_{i\in S_{X,\text{PHSP}}}\omega_{i, \text{long}} \sim S_{X,\text{long}}, \sum_{i\in S_{X,\text{PHSP}}}\omega_{i, \text{tran}} \sim S_{X,\text{tran}}, \sum_{i\in S_{X,\text{PHSP}}}\omega_{i, \text{int}} \sim S_{X,\text{int}},\\
    \end{split}
\end{equation}
where $S_{X,\text{PHSP}}$ is the shape generated using a phase-space distribution.
The contributions from longitudinal, transverse and interference components are proportional to $1:k^2:k\cos{\phi}$, based on Eq.~\ref{equ:angular_decay_rate:differential_decay_rate}. This feature allows $k$ and $|\phi|$ to be measured using a binned maximum-likelihood fit to the $m(\Dzb\Km\Kp)$ distribution in data. 
The description of the $m(\Dzb\Km)$ distribution is similar to that of $m(\Dzb\Km\Kp)$, except that there is no interference contribution.
As the correlation between the discriminating variable, $m(\Dzb\Km)$, and the control variable, $m(\Dzb\Km\Kp)$ is small, the $sFit$ technique \cite{Xie:2009rka} is used to subtract non-\Dsom decay products before fitting to $m(\Dzb\Km\Kp)$. Details of the main physical channels involved in the mass fit are described below.

\begin{flushleft}
    \bm{${\decay{\Dsom}{\Dstarzb\Km} \textbf{ decay}}$}
\end{flushleft}
As the interference term is null in the $m(\Dzb\Km)$ distribution, the partially reconstructed \decay{\Dsom}{\Dstarzb\Km} (photon/\piz chain) signals are modelled using the sum of longitudinal and transverse components. 
The efficiency-corrected yield ratio of the photon chain to the \piz chain is fixed to the branching fraction ratio between the $\decay{\Dstarzb}{\Dzb\gamma}$ and the $\decay{\Dstarzb}{\Dzb\piz}$ decays~\cite{PDG2022}
\begin{equation}
    \label{equ:mass_fit:DsoK_yield_ratio}
    \frac{N_\gamma(\Dsom)/\epsilon_\gamma(\Dsom)}{N_{\piz}(\Dsom)/\epsilon_\piz(\Dsom)}=
    \frac{\BR(\Dstarz\to\Dz\gamma)}{\BR(\Dstarz\to\Dz\piz)} = 1.83\pm0.07,
\end{equation}
where $N_\gamma(\Dsom)$ and $\epsilon_\gamma(\Dsom)$ are the signal yield and efficiency of the photon chain, respectively, while the corresponding parameters for the $\piz$ chain are $N_\piz(\Dsom)$ and $\epsilon_\piz(\Dsom)$.
The sum of yields of these two chains is a free parameter in the fit.
The probability density function of the photon/$\piz$ chain can be expressed as
\begin{equation}
    f^{\text{photon}/\piz}(m(\Dzb\Km)) = 
    f_{\text{long}}^{\text{photon}/\piz}(m(\Dzb\Km)) + k^2 f_{\text{tran}}^{\text{photon}/\piz}(m(\Dzb\Km))
\end{equation}
where $f_{\text{long}}^{\text{photon}/\piz}(m(\Dzb\Km))$ and $f_{\text{tran}}^{\text{photon}/\piz}(m(\Dzb\Km))$ correspond to the longitudinal and transverse components, respectively.

\begin{flushleft}
    \bm{${\BdsDsoK \textbf{ decay}}$}
\end{flushleft}
The partially reconstructed \BdsDsoK (photon/\piz chain) signals are modelled using the sum of longitudinal,  transverse and interference components, where the two parameters $k$ and $|\phi|$, determining the relative proportion of these components, are free parameters in the fit. 
The ratio between the photon and \piz chain yields is constrained similarly as when fitting the $m(\Dzb\Km)$ distribution using Eq.~\ref{equ:mass_fit:DsoK_yield_ratio}. The sum of yields of these two chains for each decay channel is a free parameter in the fit.
The probability density function of the photon/$\piz$ chain can be expressed as
\begin{align}
    &f^{\text{photon}/\piz}(m(\Dzb\Km\Kp)) = \\\notag
    &f_{\text{long}}^{\text{photon}/\piz}(m(\Dzb\Km\Kp)) + k^2 f_{\text{tran}}^{\text{photon}/\piz}(m(\Dzb\Km\Kp))\\\notag
    &+ k\cos(\phi)f_{\text{int}}^{\text{photon}/\piz}(m(\Dzb\Km\Kp)),
\end{align}
where $f_{\text{long}}^{\text{photon}/\piz}(m(\Dzb\Km\Kp))$, $f_{\text{tran}}^{\text{photon}/\piz}(m(\Dzb\Km\Kp))$ and $f_{\text{int}}^{\text{photon}/\piz}(m(\Dzb\Km\Kp))$ correspond to the longitudinal, transverse and interference components obtained from simulation, respectively.

\begin{flushleft}
    \bm{${\decay{\Bs}{\Dsom\pip} \textbf{ decay}}$}
\end{flushleft}
Although stringent PID criteria are applied to the kaon that does not decay from the \Dsom meson, there is still a non-negligible contribution from the \decay{\Bs}{\Dsom\pip} decay in the selected dataset. This component is modelled in the same way as the \BdsDsoK signal channels and is obtained from simulated samples, taking into account the dependence on the $k$ and $\phi$ parameters. 

\begin{flushleft}
    \bm{${\decay{\Bs}{\Dsomp\Kstarpm} \textbf{ decay}}$}
\end{flushleft}
At low $m(\Dzb\Km\Kp)$ values, there are contributions from decays which have more than one particle not reconstructed. For example, in \decay{\Bs}{\Dsomp\Kstarpm} decays, neutral particles from \Dstarzb and \Kstarp decays are not considered in the reconstruction. 
To model this contribution, simulated samples are studied. The dominant contribution of this background lies in the region with $m$(\Dzb\Km\Kp)  lower than $5000 \mevcc$ and only the right tail is present in the fitting region.
The decays \decay{\Bz}{\Dsom\rhop}, \decay{\Bm}{\Dsom\Kstarz} and its isospin partner \decay{\Bz}{\Dsomp\Kstarpm} also belong to this type of background. 
It is hard to distinguish these contributions within the current fitting range in $m(\Dzb\Km\Kp)$. Therefore in the default fit only the \decay{\Bs}{\Dsomp\Kstarpm} decay is taken into consideration. The \decay{\Bz}{\Dsom\rhop} and \decay{\Bz}{\Dsomp\Kstarpm} decay channels are considered when determining the systematic uncertainty on $k$, $|\phi|$ and $\mathcal{R}$. 
\vspace{1\baselineskip}

\begin{figure}[!tb]
    \centering
    \includegraphics[width=0.49\textwidth]{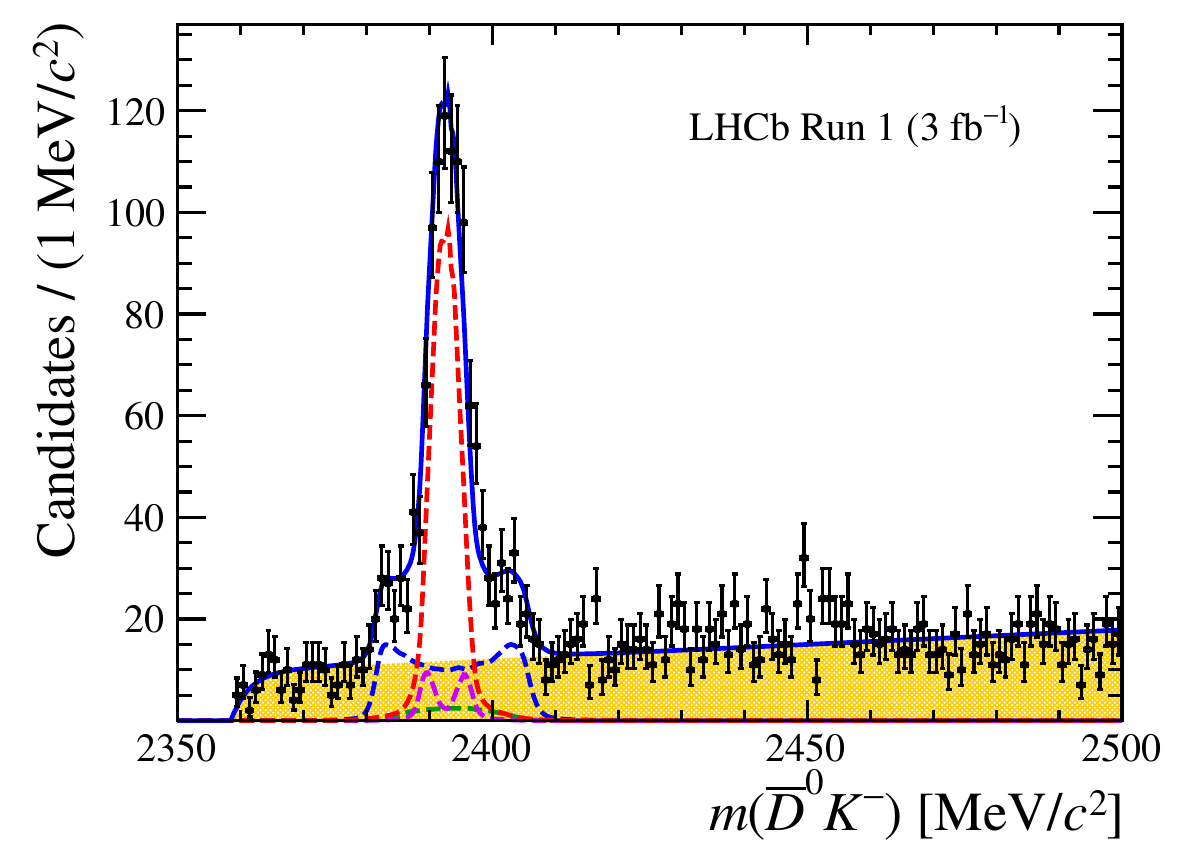}
    \includegraphics[width=0.49\textwidth]{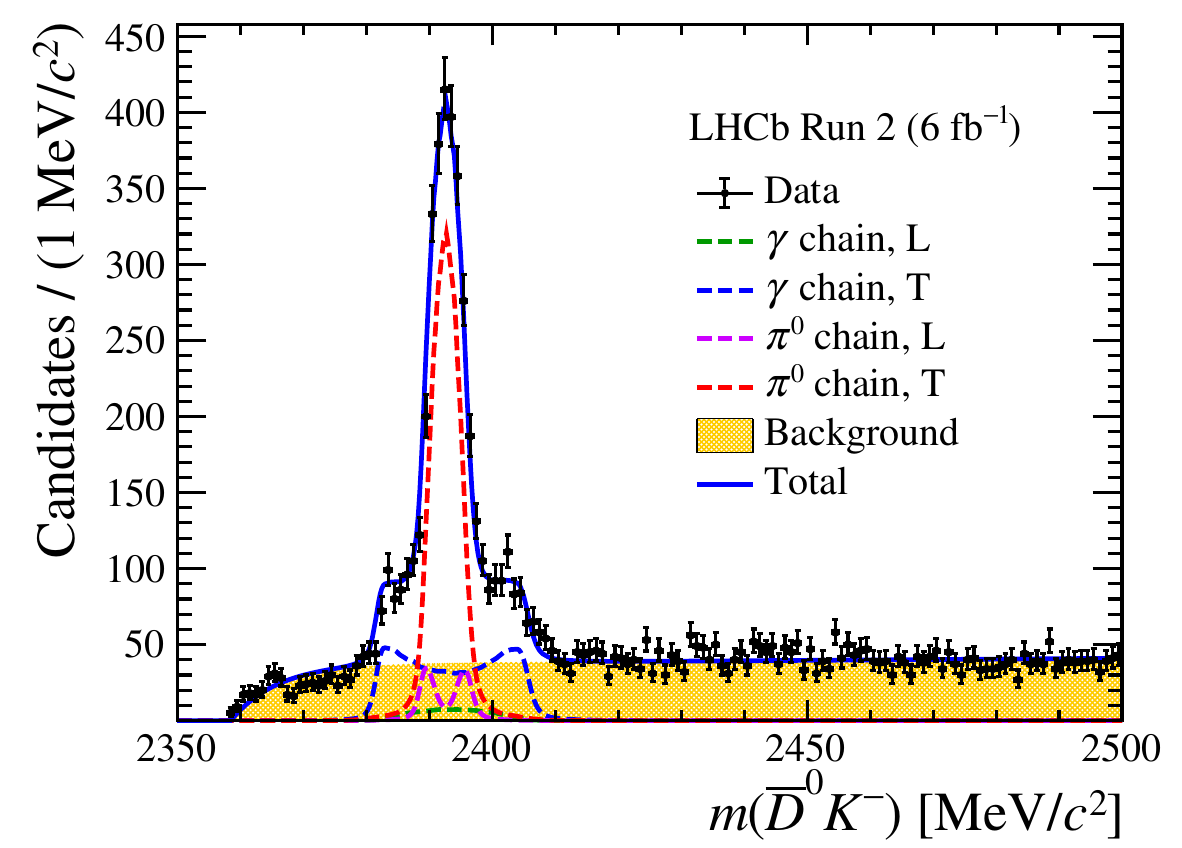}
    \caption{Mass distribution $m(\Dzb\Km)$ for selected candidates in (left) Run 1 and (right) Run 2 data, with the fit overlaid. The data is shown as black solid dots, while the blue solid line shows the results of fit. The green dashed line represents the photon chain with longitudinal polarised decay, the blue dashed line represents the photon chain with transverse polarised decay, the violet and red dashed lines describe the \piz chain with longitudinal and transverse polarisation, respectively, and the orange filled histogram represents the background.}
    \label{fig:mass_fit:DK_mass_distribution}
\end{figure}

\begin{figure}[!tb]
    \centering
    \includegraphics[width=0.49\textwidth]{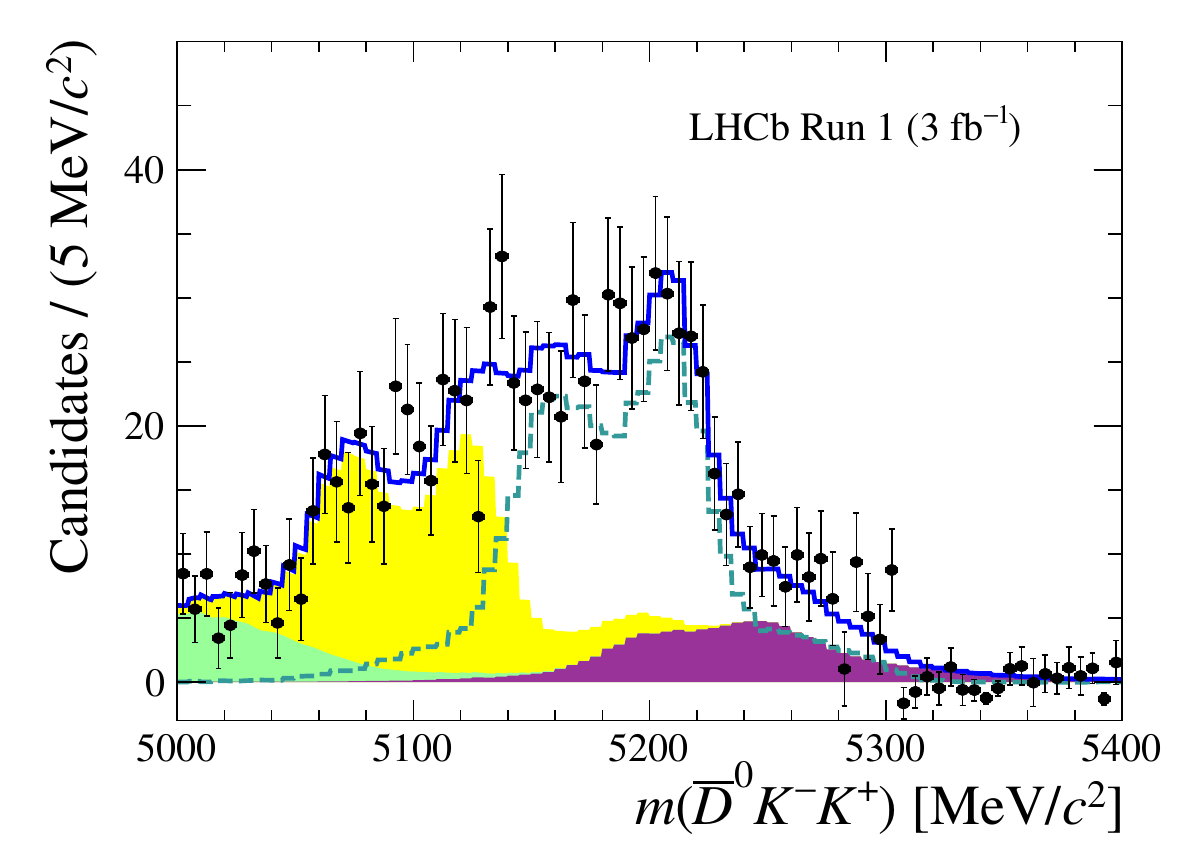}
    \includegraphics[width=0.49\textwidth]{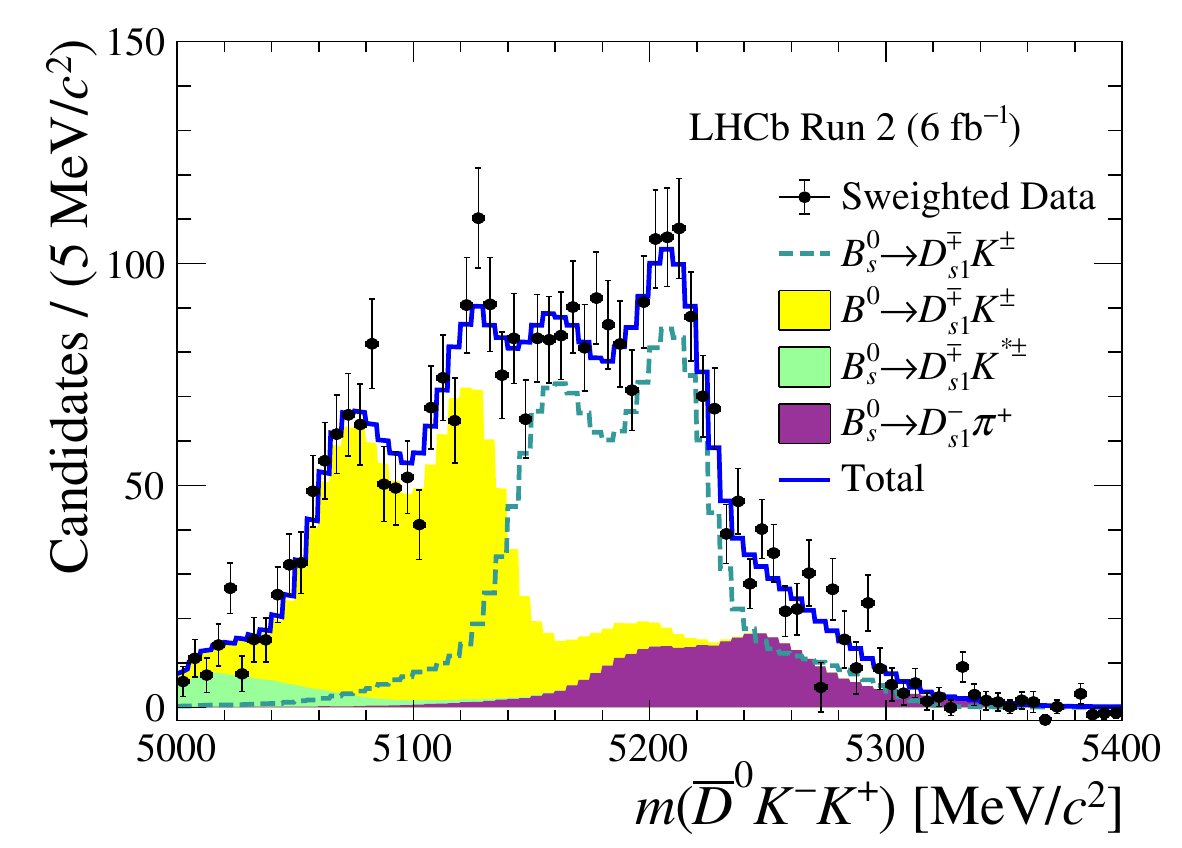}
    \caption{Mass distribution $m(\Dzb\Km\Kp)$ for selected weighted candidates in (left) Run 1 and (right) Run 2 data, with the fit overlaid. The cyan dashed line represents the ${\decay{\Bs}{\Dsomp\Kpm}}$ decay channel, the yellow filled histogram represents the ${\decay{\Bd}{\Dsomp\Kpm}}$ decay channel, the green filled histogram represents the ${\decay{\Bs}{\Dsomp\Kstarpm}}$ channel and the purple filled histogram represents the ${\decay{\Bs}{\Dsom\pip}}$ decay channel.}
    \label{fig:mass_fit:DsoK_mass_distribution}
\end{figure}

The $m(\Dzb\Km)$ mass distribution with fit projections overlaid are shown in Fig.~\ref{fig:mass_fit:DK_mass_distribution}.
The $sFit$ technique \cite{Xie:2009rka} is applied to subtract non-\Dsom decay products, when fitting the $m(\Dzb\Km\Kp)$ distribution. 
Figure~\ref{fig:mass_fit:DsoK_mass_distribution} shows the $m(\Dzb\Km\Kp)$ mass distribution with the fit results overlaid. A simultaneous fit to the two data-taking periods, Run 1 and Run 2, sharing the same $k$ and $|\phi|$ parameters, is performed. 
The distribution is dominated by two broad structures due to \decay{B_{(s)}^0}{\Dsomp\Kpm} decays. 
The combinatorial background in the $m(\Dzb\Km\Kp)$ is negligible and not considered.
To estimate the possible bias on the fitted parameters due to the $sFit$ technique \cite{Xie:2009rka}, a set of pseudoexperiment samples are generated where the correlation between these two mass variables, $m(\Dzb\Km)$ and $m(\Dzb\Km\Kp)$, is taken into account, obtained from full simulation samples which preserve correlations. 
Small biases and underestimation of uncertainties on the fitted parameters are found and corrected for.
The yields of photon/\piz chains for signal channels ${\BdsDsoK}$ and helicity-related parameters, $k$ and $|\phi|$, are summarised in Table~\ref{tab:mass_fit:DsoK_fit_result_after_cor}. The correlation coefficients between the fitted parameters are given in the Table~\ref{tab:correlation_fitted_parameters}. The significance is evaluated with a likelihood-based test, in which the likelihood distribution of the background-only hypothesis is obtained using pseudoexperiments~\cite{Cowan:2010js} taking into consideration both statistical and systematic uncertainties.
Significances of 17.9$\,\sigma$ and 12.2$\,\sigma$ are obtained for the modes $\Bs\to\Dsomp\Kpm$ and $\Bz\to\Dsomp\Kpm$, respectively.

\begin{table}[!tb]
    \centering
    \caption{
    The result of fitted parameters in the $m(\Dzb\Km\Kp)$ fit. The uncertainties are statistical.
    }
    \label{tab:mass_fit:DsoK_fit_result_after_cor}    
    \begin{tabular}{c|c|cc}
        \toprule[1pt]
        \multicolumn{2}{c|}{parameters} & Run 1 & Run 2\\
        \midrule[0.5pt]
        \multirow{2}{*}{$N(\decay{\Bs}{\Dsomp \Kpm})$} & photon chain & $154\pm13$ & $493\pm25$\\
        & \piz chain & $335\pm28$ & $1071\pm55$\\\hline
        \multirow{2}{*}{$N(\decay{\Bz}{\Dsomp \Kpm})$} & photon chain & $95\pm8$ & $374\pm15$\\
        & \piz chain & $226\pm20$ & $856\pm35$\\\hline
        \multicolumn{2}{c|}{$k$} & \multicolumn{2}{c}{\ResultKNoSys}\\
        \multicolumn{2}{c|}{$|\phi|$} & \multicolumn{2}{c}{\ResultPsiNoSys~rad}\\
        \bottomrule[1pt]
    \end{tabular}
\end{table}

\begin{table}[!tb]
    \centering
    \caption{Correlation coefficients between fitted parameters from the $m(\Dzb\Km\Kp)$ fit. $N_{\Bd}$ is the sum of the yields from photon and $\piz$ chains for \decay{\Bd}{\Dsomp\Kpm}. $N_{\Bs}$ is the sum of the yields from photon and $\piz$ chains for \decay{\Bs}{\Dsomp\Kpm}.}
    \label{tab:correlation_fitted_parameters}
    \begin{tabular}{c| *{2}{d{3.3}} | *{2}{d{3.3}} | *{1}{d{3.3}} | *{1}{d{3.3}}}
    \toprule[1pt]
         & \mc{$N_{\Bd}$(Run 1)} & \mc{$N_{\Bd}$(Run 2)} & \mc{$N_{\Bs}$(Run 1)} & \mc{$N_{\Bs}$(Run 2)} & \mc{$k$} & \mc{$|\phi|$}\\
         \midrule[0.5pt]
         $N_{\Bd}$(Run 1)& 1.00& 0.00& -0.29& 0.00& 0.00& 0.00\\
         $N_{\Bd}$(Run 2)& 0.00& 1.00& 0.00& -0.28& 0.01& 0.01\\
         \midrule[0.5pt]
         $N_{\Bs}$(Run 1)& -0.29& 0.00& 1.00& 0.00& 0.01& 0.06\\
         $N_{\Bs}$(Run 2)& 0.00& -0.28& 0.00& 1.00& 0.03& 0.06\\\hline
         $k$& 0.00& 0.01& 0.01& 0.03& 1.0& -0.26\\
         \midrule[0.5pt]
         $|\phi|$& 0.00& 0.01& 0.06& 0.06& -0.26& 1.00\\
        \bottomrule[1pt]
    \end{tabular}

\end{table}

\begin{figure}
    \centering
    \includegraphics[width=0.49\textwidth]{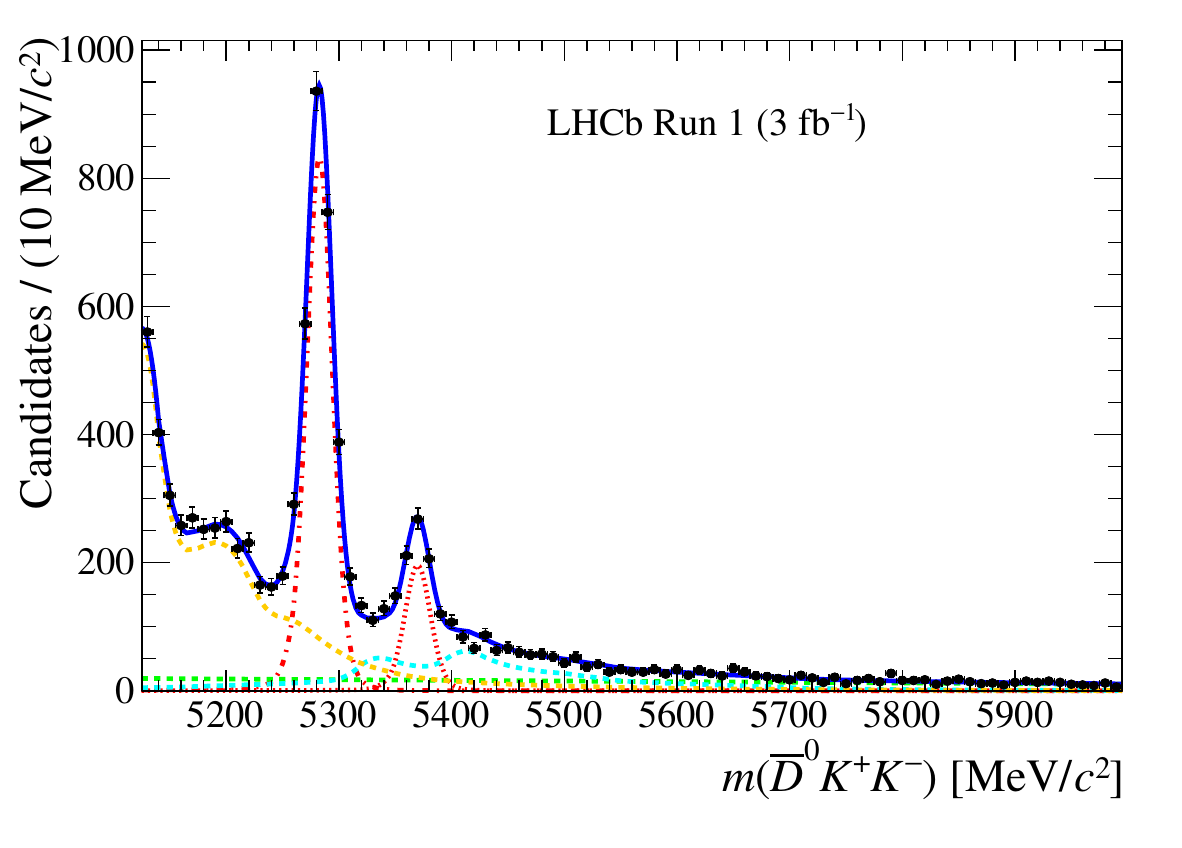}
    \includegraphics[width=0.49\textwidth]{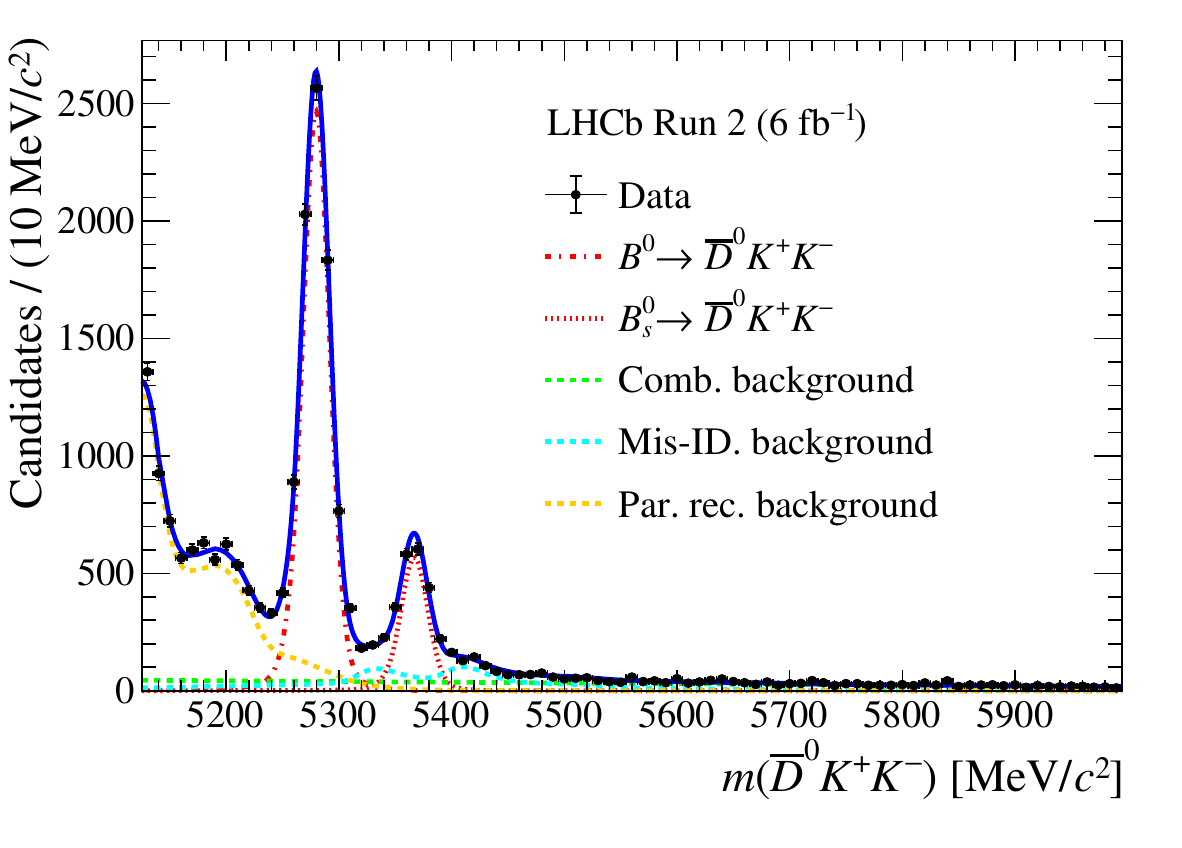}
    \caption{Mass distribution $m(\Dzb\Kp\Km)$ for selected candidates in the normalisation channel in (left) Run 1 and (right) Run 2 data, with the fit overlaid. The red dot-dashed line represents the \decay{\Bz}{\Dzb\Kp\Km} channel, the red dotted line represents the \decay{\Bs}{\Dzb\Kp\Km} channel, the green dashed line represents the combinatorial background, 
    the cyan dashed line represents the mis-identified backgrounds, which includes the \decay{\B_{(s)}^0}{\Dz\kaon\pion}, \decay{\Lb}{\Dz\proton\kaon}, \decay{\Lb}{\Dz\proton\pion} and \decay{\Lb}{\Dz\proton\pion} channels, the yellow dashed line represents the partially reconstructed background, which includes the \decay{\B^0_{(s)}}{\Dstarz\kaon\kaon} and \decay{\B^0_{(s)}}{\Dstarz\kaon\pion} channels.}
    \label{fig:mass_fit:DKK_mass_distribution_control_channel}
\end{figure}

For the normalisation channel, the selected \BdDKK candidates consist of signal and various background contributions: combinatorial, misidentified, and partially reconstructed \bquark-hadron decays. Contributions from partially reconstructed decays are reduced by the lower bounds on the mass region used in the fit. Sources of misidentified backgrounds are investigated using simulated samples. Most of the potential sources of background are found to have broad mass distributions, \eg ${\decay{\Lbbar}{\Dzb\antiproton\Kp}}$, ${\decay{\Lbbar}{\Dzb\antiproton\pip}}$ and ${\decay{\Xibbar}{\Dzb\antiproton\Kp}}$ decays. But they have obvious peaks in the \Dzb\antiproton\Kp or \Dzb\antiproton\pip mass distribution and hence their yields are determined precisely from data.
The ${\BdDKK}$ and ${\decay{\Bs}{\Dzb\Kp\Km}}$ components are described using the sum of two Crystal Ball distributions~\cite{Skwarnicki:1986xj}. The Crystal Ball distributions for both decays share common resolution and tail parameters.
The tail parameters are constrained from fits to simulation. The values of the mass and its resolution are free parameters, but the mass difference between \Bz and \Bs is fixed to the value, $\Delta m_\B = 87.45\pm0.45\mevcc$, measured by \lhcb \cite{LHCb-PAPER-2015-010}. The combinatorial background is modelled using a first-order Chebyshev polynomial, where the yield and the shape parameters of this contribution vary freely. The signal yields are obtained from unbinned extended maximum-likelihood fits to data. 
The invariant mass distributions of the normalisation channel are shown in Fig.~\ref{fig:mass_fit:DKK_mass_distribution_control_channel}. The yields are $2530\pm80$ and $7860\pm130$ for Run 1 and Run 2 data, respectively.

\section{Efficiency}
\label{sec:efficiency}

The total efficiency is factorised into the product of the detector
acceptance, reconstruction and selection efficiency, particle identification efficiency and trigger efficiency. 
The detector acceptance is the fraction of simulated decays reconstructed within the \lhcb detector. 
Selection efficiency includes the effect of the software selection in the trigger system, the initial selection, the MLP selection, and the reconstruction of the charged tracks. The efficiency in Run 2 is higher than that in Run 1 due to improvements in the reconstruction software. The PID and hardware trigger efficiencies are determined from calibration data samples where the abundant $\Dz\to\Km\pip$ sample is used~\cite{LHCb-PUB-2016-021}. The simulated PID response is corrected in order to match the data.

The simulated samples of the normalisation mode are generated uniformly over the phase space of the decay and are different from the distributions in the data. This effect is taken into account by weighting the simulation samples to match the distributions found in the data and the average efficiency is calculated using these weighted samples. 
The total efficiencies of Run 1 and Run 2 for the signal and normalisation channels are shown in Table~\ref{tab:toteff}.

\begin{table}[htbp]
\caption{Summary of total efficiencies for the signal \BdsDsoK and the normalisation mode \BdDKK, where the uncertainty is statistical.}
\begin{center}
\renewcommand\arraystretch{1.2} 
\begin{tabular}{lcc}
\hline
Mode        & \multicolumn{2}{c}{Efficiency [$10^{-4}$]} \\
    & Run~1 & Run~2 \\
\hline
\noalign{\vskip 1mm}  
\decay{\Bs}{\Dsomp\Kpm} (photon chain) & $18.70\pm0.27$ & $28.90\pm0.21$\\
\decay{\Bs}{\Dsomp\Kpm} (\piz chain) & $22.25\pm0.30$ & $34.37\pm0.26$\\
\decay{\Bd}{\Dsomp\Kpm} (photon chain) & $17.17\pm0.25$ & $25.80\pm0.20$\\
\decay{\Bd}{\Dsomp\Kpm} (\piz chain) & $22.09\pm0.30$ & $32.21\pm0.22$\\
$\BdDKK$& $13.87\pm0.07$ & $15.81\pm0.09$\\
\hline
\end{tabular} \end{center}
\label{tab:toteff}
\end{table}

\section{Systematic uncertainties}
\label{sec:systematic_uncertainty}
Systematic uncertainties on the branching fraction ratios, $\calR(\BdsDsoK)$, and the helicity-related parameters, $k$ and $|\phi|$, arise from the mass fits for the signal and normalisation channels, the limited sizes of the simulated samples, the uncertainties on the efficiency ratio corrections, and the branching fraction ratios ${\BF(\decay{\Dstarzb}{\Dzb\piz})/\BF(\decay{\Dstarzb}{\Dzb\gamma})}$. The uncertainties are given in \% relative to the measured value and are listed in Table~\ref{tab:systematic_uncertainty:summary}.

\begin{table}[!b]
    \centering
    \caption{Relative systematic uncertainties on the measurement of branching fraction ratios, $\calR(\BdsDsoK)$, and the helicity-related parameters, $k$ and $|\phi|$.}
    \label{tab:systematic_uncertainty:summary}
    \scalebox{0.9}{
    \begin{tabular}{c | cc | cc | c | c}
        \toprule[1pt] 
        \multirow{2}{2em}{Source} & \multicolumn{2}{c|}{$\calR(\decay{\Bs}{\Dsomp\Kpm})$} & \multicolumn{2}{c|}{$\calR(\decay{\Bz}{\Dsomp\Kpm})$} & $k$ & $|\phi|$ \\
        & Run 1 & Run 2 & Run 1 & Run 2 & & \\
        \midrule[0.5pt]
        \small{Signal channel fit}& 3.9\% & 2.9\% & 15.7\% & 4.2\% & 3.0\% & 4.9\%\\
        \small{Normalisation channel fit}& 1.7\% & 1.7\% & 1.7\% & 1.7\% & - & -\\
        Simulated sample sizes & 2.1\% & 1.0\% & 0.5\% & 0.3\% &  0.5\% & 3.6\% \\
        Efficiency correction & 1.2\% & 2.2\% & 2.8\% & 1.8\% & 0.7\% & 0.2\% \\
        Branching fraction ratio & 2.9\% & 2.7\% & 2.9\% & 3.1\% & 0.1\% & 0.3\%\\
        \midrule[0.5pt]
        Total &5.7\% & 5.0\% & 16.3\% & 5.8\% &3.1\%& 6.1\%\\
        \midrule[0.5pt]
        External input&\multicolumn{6}{c}{}\\
        \midrule[0.5pt]
        $f_s/f_d$ & 3.1\% & 3.1\% & - & - & - & -\\
        \bottomrule[1pt]
    \end{tabular}
    }
\end{table}

\begin{flushleft}
    \textbf{Signal channel fit}
\end{flushleft}
The choice of PDFs used to model the signal and background components in the $m(\Dzb\Km)$ or $m(\Dzb\Km\Kp)$ fit is the main source of systematic uncertainty. Taking for example the modelling of the $m(\Dzb\Km)$ distribution of the signal channels, the non-parametric PDFs from the simulation sample are used in the default fit, then the RooHILLdini and RooHORNdini functions are used as alternative models, which are usually able to describe the shape of the mass distribution of vector decays and are defined in Ref.~\cite{LHCb-PAPER-2017-021}. 
To determine this uncertainty, the alternative fit model is used to generate a set of pseudoexperiments and the default model is used to fit them. The bias obtained from these fits is taken as the corresponding systematic uncertainty. 

The most significant contribution to this systematic uncertainty arises from the physical background description in the fit of the $m(\Dzb\Km\Kp)$ distribution.
In the lower side of the fit region, the primary contribution comes from $\decay{\Bz}{\Dsomp \Kpm}$ decays. In addition, there is a need for contributions from other components, such as $\decay{\Bs}{\Dsomp \Kstarpm}$ $\decay{\Bz}{\Dsom\rhop}$ or $\decay{\Bz}{\Dsomp \Kstarpm}$ decays, which are discussed in Sec.~\ref{sec:mass_fit}. According to simulation, all of them are likely to enter the fit region, albeit with different slopes. Attempts to include each component independently into the fit model have been made, and the case resulting in the largest change in yield is used to calculate the systematic uncertainty. Since the lower side of the fit region is primarily used to determine the yields of $\decay{\Bz}{\Dsomp \Kpm}$ decays, the uncertainties for results from $\Bz$ decays are larger than those from $\Bs$ decays.

In addition, the $sFit$ method is used to subtract background contributions when fitting the $m(\Dzb\Km\Kp)$ distribution. It assumes no correlation between the discriminating variable $m(\Dzb\Km)$ and the control variable $m(\Dzb\Km\Kp)$. To estimate the corresponding systematic uncertainty, the fit procedure is performed with 1,500 pseudoexperiments samples obtained from full simulation which preserves correlations. The bias obtained from these fits is assigned as the systematic uncertainty. 
The systematic uncertainty from the fit model for the signal channel is considered fully correlated between Run 1 and Run 2 data samples.

\begin{flushleft}
    \textbf{Normalisation channel fit}
\end{flushleft}
The signal model for the normalisation channel is the sum of two Crystal Ball distributions with tail parameters determined from simulation. To determine the related systematic uncertainties, the fit to the $m(\Dzb\Kp\Km)$ distribution is performed many times with the parameters randomly varied within their uncertainties according to Gaussian distributions. The width of the distribution of the normalised residuals of the yields is assigned as a systematic uncertainty.
The background model is changed to an exponential function to estimate the systematic uncertainty and the difference with respect to the default model is assigned as a systematic uncertainty. This source of systematic uncertainty only affects the \BdDKK yield and does not contribute to the helicity-related parameters, $k$ and $|\phi|$.
This source of systematic uncertainty is considered as fully correlated between the Run 1 and Run 2 data samples.
\begin{flushleft}
    \textbf{Simulated sample size}
\end{flushleft}
In the $m(\Dzb\Km\Kp)$ fit, the efficiency-corrected ratios of yields between the photon \textrm{chain} and the \piz \textrm{chain} are fixed. Statistical uncertainties of efficiencies due to limited simulated sample size, therefore, affect the ratios and are taken as a source of systematic uncertainty. To determine this systematic uncertainty, the $m(\Dzb\Kp\Km)$ fit to data is performed 1,000 times with efficiencies Gaussian constrained. The width of the distribution of the fitted yields is taken as a systematic uncertainty. This source of systematic uncertainty is considered as uncorrelated between the Run 1 and Run 2 samples due to its statistical nature.

\begin{flushleft}
    \textbf{Efficiency correction}
\end{flushleft}
Differences between data and simulation are corrected using data and calibration samples.  
The systematic uncertainty due to the limited size of the calibration samples is considered. It is composed of contributions from the  PID efficiency and the hardware trigger efficiency. The systematic uncertainty is evaluated using the same method used for the simulated sample size. Almost all of this source of systematic uncertainty is considered to be uncorrelated between the Run 1 and Run 2 samples, except the one related to the detector acceptance that is considered as fully correlated. 

\begin{flushleft}
    \textbf{Branching fraction ratio}
\end{flushleft}
In the $m(\Dzb\Km\Kp)$ fit, the efficiency-corrected ratios of yields between the photon and \piz \textrm{chains} is related to the branching fraction ratio, ${\BF(\decay{\Dstarz}{\Dz\piz})/\BF(\decay{\Dstarz}{\Dz\gamma}) = 1.83\pm0.07}$. In the default fit, this branching fraction ratio is fixed to its central value. To evaluate the related systematic uncertainty, the branching fraction ratio is varied by one standard deviation. The fit is repeated with the new values of the branching fraction ratio, and the relative difference is assigned as a systematic uncertainty which is considered as fully correlated between Run 1 and Run 2 samples.

\begin{flushleft}
    \bm{$\textbf{Fragmentation fraction ratio:~} {f_s/f_d}$}
\end{flushleft}
The fragmentation fraction ratio, $f_s/f_d$, of a $b$ quark into a $\Bs$ and a $\Bz$ meson in \lhcb  proton-proton collisions~\cite{LHCb-PAPER-2020-046} is an important source of systematic uncertainty for the $\decay{\Bs}{\Dsomp\Kpm}$ decay and depends on the centre-of-mass energy for the collected data. The values of this quantity are $0.2387\pm0.0075$ and $0.2539\pm0.0079$ for the Runs 1 and 2, respectively. 
The uncertainty from $f_s/f_d$ is propagated as a systematic uncertainty, which is considered as uncorrelated between the Run 1 and Run 2 samples.

\vspace{1\baselineskip}
For each data taking period, the total systematic uncertainty is obtained as the quadratic sum of all contributions, except the systematic uncertainty from $f_s/f_d$, which is listed separately, and shown in Table  \ref{tab:systematic_uncertainty:summary}.

\section{Results}
\label{sec:results}
The \BdsDsoK branching fractions relative to that of the \BdDKK decay channel are defined as
\begin{equation}
    \label{equ:conclusion:ana_strategy_12}
    \begin{split}
        & \calR(\decay{\Bs}{\Dsomp\Kpm}) = \\
        & \frac{ \left( \calN_{\text{cor}}\left( \decay{\Bs}{\Dsomp \Kpm}, \gamma \textrm{~chain}\right) + \calN_{\text{cor}}\left( \decay{\Bs}{\Dsomp \Kpm}, \piz \textrm{~chain}\right) \right)} {\calN_{\text{cor}}\left( \decay{\Bz}{\Dzb \Kp \Km}\right) \times f_\squark/f_\dquark},\\
        &\calR(\decay{\Bz}{\Dsomp \Kpm}) = \\
        & \frac{ \left( \calN_{\text{cor}}\left( \decay{\Bz}{\Dsomp \Kpm}, \gamma \textrm{~chain}\right) + \calN_{\text{cor}}\left( \decay{\Bz}{\Dsomp \Kpm}, \piz \textrm{~chain}\right) \right)} {\calN_{\text{cor}}\left( \decay{\Bz}{\Dzb \Kp \Km}\right)},\\
    \end{split}
\end{equation}
where ${f_\squark}$ (${f_\dquark}$) is the fragmentation fraction of a $b$ quark into a $\Bs$ ($\Bd$) meson, 
and $\calN_{\text{cor}}(X)$ is the efficiency-corrected yield, $N(X)/\epsilon(X)$, of the corresponding channel $X$. 
Using the corrected yields determined in the previous sections  and the ratio $f_s/f_d$ , the branching fraction ratios,  \mbox{\calR(\BdsDsoK)}, are determined to be
\begin{align*}
    &\calR(\decay{\Bs}{\Dsomp \Kpm}) = \ResultRelaBRBsDsoKDstK,\\
    &\calR(\decay{\Bz}{\Dsomp \Kpm}) = \ResultRelaBRBzDsoKDstK,
\end{align*}
where the first uncertainty is statistical, the second systematic, and the third due to the uncertainty of $f_s/f_d$. 
Using $\BR(\BdDKK)=(6.1\pm0.4\pm0.3\pm0.3)\times10^{-5}$~\cite{LHCb-PAPER-2018-014}, the absolute branching fractions are calculated to be 
\begin{equation*}
    \begin{split}
        \BF(\decay{\Bs}{\Dsomp\Kpm})\times& \BR(\decay{\Dsom}{\Dstarzb\Km}) =\\
        &\ResultBRBsDsoKDstK,\\
        \BF(\decay{\Bz}{\Dsomp\Kpm})\times &\BR(\decay{\Dsom}{\Dstarzb\Km})=\\
        &\ResultBRBzDsoKDstK,
    \end{split}
\end{equation*}
where the third uncertainties are due to the uncertainty of $\BR(\BdDKK)$, and the fourth from the $f_s/f_d$ ratio.

The helicity-related parameters $k$ and $|\phi|$, used to describe the ratio and phase difference between the helicity amplitudes $H_+$ and $H_0$, are determined to be
\begin{gather*} 
   k = \ResultK,\quad |\phi| = \ResultPsi~\rad,
\end{gather*}
where the first uncertainty is statistical and the second systematic.

The helicity amplitudes are related to the $LS$ couplings ($B_{L,S}$) using the Clebsch-Gordan coefficients,
where $L$ is the orbital angular momentum in a two body decay $\decay{A}{B+C}$, and $S$ is the total spin of the secondary particles, ${\Vec{S} = \Vec{J_{B}} + \Vec{J_{C}}}$ ${(| J_{B}-J_{C}|\leq S \leq J_{B}+J_{C})}$, where $J_{B}$ and $J_{C}$ are the spins of $B$ and $C$ particles. 
This leads to the following relations:
\begin{gather*}
    H_{+} = \frac{\sqrt{3}}{3}B_{0, 1} + \frac{\sqrt{6}}{6} B_{2, 1}=H_{-},\\
    H_{0} = \frac{\sqrt{3}}{3}B_{0, 1} - \frac{\sqrt{6}}{3} B_{2, 1},
\end{gather*}
where $B_{0, 1}$ corresponds to the $S$-wave and $B_{2, 1}$ to the $D$-wave. 
The amplitude ratio between $S$- and $D$-waves, $B_{0,1}/B_{2,1}\equiv A e^{iB}$, are determined to be 
\begin{gather*}
    A = \ResultA,\quad |B| = \ResultB~\rad,
\end{gather*}
where the first uncertainty is statistical and the second systematic. The fraction of $S$-wave component in \decay{\Dsop}{\Dstar^0\Kp} is calculated to be $(55\pm7\pm3)$\%, consistent with the results from its isospin partner \decay{\Dsop}{\Dstarp\Kz}, in which the $S$-wave fraction is $(72\pm5\pm1)$\% \cite{Belle:2007kff}.

\section{Conclusion}
\label{sec:conclusion}
In summary, the decays \decay{\B^0_{(s)}}{\Dsomp\Kpm} are observed for the first time and their branching fractions are measured using a data sample corresponding to an integrated luminosity of $9\invfb$ of $\proton\proton$ collisions collected by the \lhcb experiment with the significance for the \Bs and \Bd signals larger than $10\,\sigma$.
The fraction of the $S$-wave component in the \decay{\Dsop}{\Dstarz\Kp} decay is determined to be $(55\pm7\pm3)$\%, which is comparable to that of its isospin partner \decay{\Dsop}{\Dstarp\Kz}~\cite{Belle:2007kff}.
The observation of the ${\decay{\Bs}{\Dsomp\Kpm}}$ decay channel is the first step towards the CKM $\gamma$ angle extraction using this channel.
By the end of the \lhcb Run 3 data taking period, the signal yields will allow us to measure the $\gamma$ angle with a precision similar to that of \decay{\Bs}{\Dsmp\Kpm} decay in Run 1~\cite{LHCb-PAPER-2017-047}, which could provide useful information as a reference value to be compared with theory predictions.
In addition, the results of the branching fractions of the \decay{\Bds}{\Dsomp\Kpm} decay channels can be compared with theoretical predictions with QCD factorisation to shed light on whether or not there is physics beyond the Standard Model in the \decay{b}{c\bar{u}s} and \decay{b}{\bar{u}cs} processes.

\section*{Acknowledgements}
%
%
\noindent We express our gratitude to our colleagues in the CERN
accelerator departments for the excellent performance of the LHC. We
thank the technical and administrative staff at the LHCb
institutes.
We acknowledge support from CERN and from the national agencies:
CAPES, CNPq, FAPERJ and FINEP (Brazil); 
MOST and NSFC (China); 
CNRS/IN2P3 (France); 
BMBF, DFG and MPG (Germany); 
INFN (Italy); 
NWO (Netherlands); 
MNiSW and NCN (Poland); 
MCID/IFA (Romania); 
MICINN (Spain); 
SNSF and SER (Switzerland); 
NASU (Ukraine); 
STFC (United Kingdom); 
DOE NP and NSF (USA).
We acknowledge the computing resources that are provided by CERN, IN2P3
(France), KIT and DESY (Germany), INFN (Italy), SURF (Netherlands),
PIC (Spain), GridPP (United Kingdom), 
CSCS (Switzerland), IFIN-HH (Romania), CBPF (Brazil),
Polish WLCG  (Poland) and NERSC (USA).
We are indebted to the communities behind the multiple open-source
software packages on which we depend.
Individual groups or members have received support from
ARC and ARDC (Australia);
Minciencias (Colombia);
AvH Foundation (Germany);
EPLANET, Marie Sk\l{}odowska-Curie Actions, ERC and NextGenerationEU (European Union);
A*MIDEX, ANR, IPhU and Labex P2IO, and R\'{e}gion Auvergne-Rh\^{o}ne-Alpes (France);
Key Research Program of Frontier Sciences of CAS, CAS PIFI, CAS CCEPP, 
Fundamental Research Funds for the Central Universities, 
and Sci. \& Tech. Program of Guangzhou (China);
GVA, XuntaGal, GENCAT, Inditex, InTalent and Prog.~Atracci\'on Talento, CM (Spain);
SRC (Sweden);
the Leverhulme Trust, the Royal Society
 and UKRI (United Kingdom).

\addcontentsline{toc}{section}{References}
\bibliographystyle{LHCb}
\bibliography{main,standard,LHCb-PAPER,LHCb-CONF,LHCb-DP,LHCb-TDR}

\ifx\mcitethebibliography\mciteundefinedmacro
\PackageError{LHCb.bst}{mciteplus.sty has not been loaded}
{This bibstyle requires the use of the mciteplus package.}\fi
\providecommand{\href}[2]{#2}
\begin{mcitethebibliography}{10}
\mciteSetBstSublistMode{n}
\mciteSetBstMaxWidthForm{subitem}{\alph{mcitesubitemcount})}
\mciteSetBstSublistLabelBeginEnd{\mcitemaxwidthsubitemform\space}
{\relax}{\relax}

\bibitem{Cabibbo:1963yz}
N.~Cabibbo, \ifthenelse{\boolean{articletitles}}{\emph{{Unitary symmetry and
  leptonic decays}},
  }{}\href{https://doi.org/10.1103/PhysRevLett.10.531}{Phys.\ Rev.\ Lett.\
  \textbf{10} (1963) 531}\relax
\mciteBstWouldAddEndPuncttrue
\mciteSetBstMidEndSepPunct{\mcitedefaultmidpunct}
{\mcitedefaultendpunct}{\mcitedefaultseppunct}\relax
\EndOfBibitem
\bibitem{Kobayashi:1973fv}
M.~Kobayashi and T.~Maskawa,
  \ifthenelse{\boolean{articletitles}}{\emph{{\CP-violation in the
  renormalizable theory of weak interaction}},
  }{}\href{https://doi.org/10.1143/PTP.49.652}{Prog.\ Theor.\ Phys.\
  \textbf{49} (1973) 652}\relax
\mciteBstWouldAddEndPuncttrue
\mciteSetBstMidEndSepPunct{\mcitedefaultmidpunct}
{\mcitedefaultendpunct}{\mcitedefaultseppunct}\relax
\EndOfBibitem
\bibitem{LHCb-PAPER-2017-047}
LHCb collaboration, R.~Aaij {\em et~al.},
  \ifthenelse{\boolean{articletitles}}{\emph{{Measurement of \CP asymmetry in
  \mbox{\decay{\Bs}{D_s^\mp \Kpm}} decays}},
  }{}\href{https://doi.org/10.1007/JHEP03(2018)059}{JHEP \textbf{03} (2018)
  059}, \href{http://arxiv.org/abs/1712.07428}{{\normalfont\ttfamily
  arXiv:1712.07428}}\relax
\mciteBstWouldAddEndPuncttrue
\mciteSetBstMidEndSepPunct{\mcitedefaultmidpunct}
{\mcitedefaultendpunct}{\mcitedefaultseppunct}\relax
\EndOfBibitem
\bibitem{LHCB-PAPER-2020-030}
LHCb collaboration, R.~Aaij {\em et~al.},
  \ifthenelse{\boolean{articletitles}}{\emph{{Measurement of the CKM angle
  $\gamma$ and \mbox{\Bs-\Bsb} mixing frequency with $\Bs \to \Dsmp h^\pm \pipm
  \pimp$ decays}}, }{}\href{https://doi.org/10.1007/JHEP03(2021)137}{JHEP
  \textbf{03} (2021) 137},
  \href{http://arxiv.org/abs/2011.12041}{{\normalfont\ttfamily
  arXiv:2011.12041}}\relax
\mciteBstWouldAddEndPuncttrue
\mciteSetBstMidEndSepPunct{\mcitedefaultmidpunct}
{\mcitedefaultendpunct}{\mcitedefaultseppunct}\relax
\EndOfBibitem
\bibitem{PDG2022}
Particle Data Group, R.~L. Workman and Others,
  \ifthenelse{\boolean{articletitles}}{\emph{{Review of Particle Physics}},
  }{}\href{https://doi.org/10.1093/ptep/ptac097}{PTEP \textbf{2022} (2022)
  083C01}\relax
\mciteBstWouldAddEndPuncttrue
\mciteSetBstMidEndSepPunct{\mcitedefaultmidpunct}
{\mcitedefaultendpunct}{\mcitedefaultseppunct}\relax
\EndOfBibitem
\bibitem{Fleischer:2010ca}
R.~Fleischer, N.~Serra, and N.~Tuning,
  \ifthenelse{\boolean{articletitles}}{\emph{Tests of factorization and {SU(3)}
  relations in {$B$} decays into heavy-light final states},
  }{}\href{https://doi.org/10.1103/PhysRevD.83.014017}{Phys.\ Rev.\
  \textbf{D83} (2011) 014017},
  \href{http://arxiv.org/abs/1012.2784}{{\normalfont\ttfamily
  arXiv:1012.2784}}\relax
\mciteBstWouldAddEndPuncttrue
\mciteSetBstMidEndSepPunct{\mcitedefaultmidpunct}
{\mcitedefaultendpunct}{\mcitedefaultseppunct}\relax
\EndOfBibitem
\bibitem{Bordone:2020gao}
M.~Bordone {\em et~al.}, \ifthenelse{\boolean{articletitles}}{\emph{{A puzzle
  in $\bar{B}_{(s)}^0 \to D_{(s)}^{(*)+} \lbrace \pi^-, K^-\rbrace$ decays and
  extraction of the $f_s/f_d$ fragmentation fraction}},
  }{}\href{https://doi.org/10.1140/epjc/s10052-020-08512-8}{Eur.\ Phys.\ J.\
  \textbf{C80} (2020) 951},
  \href{http://arxiv.org/abs/2007.10338}{{\normalfont\ttfamily
  arXiv:2007.10338}}\relax
\mciteBstWouldAddEndPuncttrue
\mciteSetBstMidEndSepPunct{\mcitedefaultmidpunct}
{\mcitedefaultendpunct}{\mcitedefaultseppunct}\relax
\EndOfBibitem
\bibitem{Cai:2021mlt}
F.-M. Cai, W.-J. Deng, X.-Q. Li, and Y.-D. Yang,
  \ifthenelse{\boolean{articletitles}}{\emph{{Probing new physics in class-I
  B-meson decays into heavy-light final states}},
  }{}\href{https://doi.org/10.1007/JHEP10(2021)235}{JHEP \textbf{10} (2021)
  235}, \href{http://arxiv.org/abs/2103.04138}{{\normalfont\ttfamily
  arXiv:2103.04138}}\relax
\mciteBstWouldAddEndPuncttrue
\mciteSetBstMidEndSepPunct{\mcitedefaultmidpunct}
{\mcitedefaultendpunct}{\mcitedefaultseppunct}\relax
\EndOfBibitem
\bibitem{Endo:2021ifc}
M.~Endo, S.~Iguro, and S.~Mishima,
  \ifthenelse{\boolean{articletitles}}{\emph{{Revisiting rescattering
  contributions to ${\overline{B}_{(s)}\to{D}_{(s)}^{\left(\ast \right)}M}$
  decays}}, }{}\href{https://doi.org/10.1007/JHEP01(2022)147}{JHEP \textbf{01}
  (2022) 147}, \href{http://arxiv.org/abs/2109.10811}{{\normalfont\ttfamily
  arXiv:2109.10811}}\relax
\mciteBstWouldAddEndPuncttrue
\mciteSetBstMidEndSepPunct{\mcitedefaultmidpunct}
{\mcitedefaultendpunct}{\mcitedefaultseppunct}\relax
\EndOfBibitem
\bibitem{Fleischer:2021cct}
R.~Fleischer and E.~Malami, \ifthenelse{\boolean{articletitles}}{\emph{{Using
  $B^0_s\to D_s^\mp K^\pm$ Decays as a Portal to New Physics}},
  }{}\href{https://doi.org/10.1103/PhysRevD.106.056004}{Phys.\ Rev.\
  \textbf{D106} (2022) 056004},
  \href{http://arxiv.org/abs/2109.04950}{{\normalfont\ttfamily
  arXiv:2109.04950}}\relax
\mciteBstWouldAddEndPuncttrue
\mciteSetBstMidEndSepPunct{\mcitedefaultmidpunct}
{\mcitedefaultendpunct}{\mcitedefaultseppunct}\relax
\EndOfBibitem
\bibitem{LHCb-PAPER-2015-051}
LHCb collaboration, R.~Aaij {\em et~al.},
  \ifthenelse{\boolean{articletitles}}{\emph{{Angular analysis of the
  \mbox{\decay{\Bz}{\Kstarz\mumu}} decay using $3\invfb$ of integrated
  luminosity}}, }{}\href{https://doi.org/10.1007/JHEP02(2016)104}{JHEP
  \textbf{02} (2016) 104},
  \href{http://arxiv.org/abs/1512.04442}{{\normalfont\ttfamily
  arXiv:1512.04442}}\relax
\mciteBstWouldAddEndPuncttrue
\mciteSetBstMidEndSepPunct{\mcitedefaultmidpunct}
{\mcitedefaultendpunct}{\mcitedefaultseppunct}\relax
\EndOfBibitem
\bibitem{LHCb-PAPER-2020-041}
LHCb collaboration, R.~Aaij {\em et~al.},
  \ifthenelse{\boolean{articletitles}}{\emph{{Angular analysis of the $B^{+}\to
  K^{\ast+}\mu^+\mu^-$ decay}},
  }{}\href{https://doi.org/10.1103/PhysRevLett.126.161802}{Phys.\ Rev.\ Lett.\
  \textbf{126} (2021) 161802},
  \href{http://arxiv.org/abs/2012.13241}{{\normalfont\ttfamily
  arXiv:2012.13241}}\relax
\mciteBstWouldAddEndPuncttrue
\mciteSetBstMidEndSepPunct{\mcitedefaultmidpunct}
{\mcitedefaultendpunct}{\mcitedefaultseppunct}\relax
\EndOfBibitem
\bibitem{LHCb-DP-2008-001}
LHCb collaboration, A.~A. Alves~Jr.\ {\em et~al.},
  \ifthenelse{\boolean{articletitles}}{\emph{{The \lhcb detector at the LHC}},
  }{}\href{https://doi.org/10.1088/1748-0221/3/08/S08005}{JINST \textbf{3}
  (2008) S08005}\relax
\mciteBstWouldAddEndPuncttrue
\mciteSetBstMidEndSepPunct{\mcitedefaultmidpunct}
{\mcitedefaultendpunct}{\mcitedefaultseppunct}\relax
\EndOfBibitem
\bibitem{LHCb-DP-2014-002}
LHCb collaboration, R.~Aaij {\em et~al.},
  \ifthenelse{\boolean{articletitles}}{\emph{{LHCb detector performance}},
  }{}\href{https://doi.org/10.1142/S0217751X15300227}{Int.\ J.\ Mod.\ Phys.\
  \textbf{A30} (2015) 1530022},
  \href{http://arxiv.org/abs/1412.6352}{{\normalfont\ttfamily
  arXiv:1412.6352}}\relax
\mciteBstWouldAddEndPuncttrue
\mciteSetBstMidEndSepPunct{\mcitedefaultmidpunct}
{\mcitedefaultendpunct}{\mcitedefaultseppunct}\relax
\EndOfBibitem
\bibitem{BBDT}
V.~V. Gligorov and M.~Williams,
  \ifthenelse{\boolean{articletitles}}{\emph{{Efficient, reliable and fast
  high-level triggering using a bonsai boosted decision tree}},
  }{}\href{https://doi.org/10.1088/1748-0221/8/02/P02013}{JINST \textbf{8}
  (2013) P02013}, \href{http://arxiv.org/abs/1210.6861}{{\normalfont\ttfamily
  arXiv:1210.6861}}\relax
\mciteBstWouldAddEndPuncttrue
\mciteSetBstMidEndSepPunct{\mcitedefaultmidpunct}
{\mcitedefaultendpunct}{\mcitedefaultseppunct}\relax
\EndOfBibitem
\bibitem{LHCb-PROC-2015-018}
T.~Likhomanenko {\em et~al.}, \ifthenelse{\boolean{articletitles}}{\emph{{LHCb
  topological trigger reoptimization}},
  }{}\href{https://doi.org/10.1088/1742-6596/664/8/082025}{J.\ Phys.\ Conf.\
  Ser.\  \textbf{664} (2015) 082025}\relax
\mciteBstWouldAddEndPuncttrue
\mciteSetBstMidEndSepPunct{\mcitedefaultmidpunct}
{\mcitedefaultendpunct}{\mcitedefaultseppunct}\relax
\EndOfBibitem
\bibitem{Sjostrand:2007gs}
T.~Sj\"{o}strand, S.~Mrenna, and P.~Skands,
  \ifthenelse{\boolean{articletitles}}{\emph{{A brief introduction to PYTHIA
  8.1}}, }{}\href{https://doi.org/10.1016/j.cpc.2008.01.036}{Comput.\ Phys.\
  Commun.\  \textbf{178} (2008) 852},
  \href{http://arxiv.org/abs/0710.3820}{{\normalfont\ttfamily
  arXiv:0710.3820}}\relax
\mciteBstWouldAddEndPuncttrue
\mciteSetBstMidEndSepPunct{\mcitedefaultmidpunct}
{\mcitedefaultendpunct}{\mcitedefaultseppunct}\relax
\EndOfBibitem
\bibitem{Sjostrand:2006za}
T.~Sj\"{o}strand, S.~Mrenna, and P.~Skands,
  \ifthenelse{\boolean{articletitles}}{\emph{{PYTHIA 6.4 physics and manual}},
  }{}\href{https://doi.org/10.1088/1126-6708/2006/05/026}{JHEP \textbf{05}
  (2006) 026}, \href{http://arxiv.org/abs/hep-ph/0603175}{{\normalfont\ttfamily
  arXiv:hep-ph/0603175}}\relax
\mciteBstWouldAddEndPuncttrue
\mciteSetBstMidEndSepPunct{\mcitedefaultmidpunct}
{\mcitedefaultendpunct}{\mcitedefaultseppunct}\relax
\EndOfBibitem
\bibitem{LHCb-PROC-2010-056}
I.~Belyaev {\em et~al.}, \ifthenelse{\boolean{articletitles}}{\emph{{Handling
  of the generation of primary events in Gauss, the LHCb simulation
  framework}}, }{}\href{https://doi.org/10.1088/1742-6596/331/3/032047}{J.\
  Phys.\ Conf.\ Ser.\  \textbf{331} (2011) 032047}\relax
\mciteBstWouldAddEndPuncttrue
\mciteSetBstMidEndSepPunct{\mcitedefaultmidpunct}
{\mcitedefaultendpunct}{\mcitedefaultseppunct}\relax
\EndOfBibitem
\bibitem{Lange:2001uf}
D.~J. Lange, \ifthenelse{\boolean{articletitles}}{\emph{{The EvtGen particle
  decay simulation package}},
  }{}\href{https://doi.org/10.1016/S0168-9002(01)00089-4}{Nucl.\ Instrum.\
  Meth.\  \textbf{A462} (2001) 152}\relax
\mciteBstWouldAddEndPuncttrue
\mciteSetBstMidEndSepPunct{\mcitedefaultmidpunct}
{\mcitedefaultendpunct}{\mcitedefaultseppunct}\relax
\EndOfBibitem
\bibitem{davidson2015photos}
N.~Davidson, T.~Przedzinski, and Z.~Was,
  \ifthenelse{\boolean{articletitles}}{\emph{{PHOTOS interface in C++:
  Technical and physics documentation}},
  }{}\href{https://doi.org/https://doi.org/10.1016/j.cpc.2015.09.013}{Comp.\
  Phys.\ Comm.\  \textbf{199} (2016) 86},
  \href{http://arxiv.org/abs/1011.0937}{{\normalfont\ttfamily
  arXiv:1011.0937}}\relax
\mciteBstWouldAddEndPuncttrue
\mciteSetBstMidEndSepPunct{\mcitedefaultmidpunct}
{\mcitedefaultendpunct}{\mcitedefaultseppunct}\relax
\EndOfBibitem
\bibitem{Allison:2006ve}
Geant4 collaboration, J.~Allison {\em et~al.},
  \ifthenelse{\boolean{articletitles}}{\emph{{Geant4 developments and
  applications}}, }{}\href{https://doi.org/10.1109/TNS.2006.869826}{IEEE
  Trans.\ Nucl.\ Sci.\  \textbf{53} (2006) 270}\relax
\mciteBstWouldAddEndPuncttrue
\mciteSetBstMidEndSepPunct{\mcitedefaultmidpunct}
{\mcitedefaultendpunct}{\mcitedefaultseppunct}\relax
\EndOfBibitem
\bibitem{Agostinelli:2002hh}
Geant4 collaboration, S.~Agostinelli {\em et~al.},
  \ifthenelse{\boolean{articletitles}}{\emph{{Geant4: A simulation toolkit}},
  }{}\href{https://doi.org/10.1016/S0168-9002(03)01368-8}{Nucl.\ Instrum.\
  Meth.\  \textbf{A506} (2003) 250}\relax
\mciteBstWouldAddEndPuncttrue
\mciteSetBstMidEndSepPunct{\mcitedefaultmidpunct}
{\mcitedefaultendpunct}{\mcitedefaultseppunct}\relax
\EndOfBibitem
\bibitem{LHCb-PROC-2011-006}
M.~Clemencic {\em et~al.}, \ifthenelse{\boolean{articletitles}}{\emph{{The
  \lhcb simulation application, Gauss: Design, evolution and experience}},
  }{}\href{https://doi.org/10.1088/1742-6596/331/3/032023}{J.\ Phys.\ Conf.\
  Ser.\  \textbf{331} (2011) 032023}\relax
\mciteBstWouldAddEndPuncttrue
\mciteSetBstMidEndSepPunct{\mcitedefaultmidpunct}
{\mcitedefaultendpunct}{\mcitedefaultseppunct}\relax
\EndOfBibitem
\bibitem{Hulsbergen:2005pu}
W.~D. Hulsbergen, \ifthenelse{\boolean{articletitles}}{\emph{{Decay chain
  fitting with a Kalman filter}},
  }{}\href{https://doi.org/10.1016/j.nima.2005.06.078}{Nucl.\ Instrum.\ Meth.\
  \textbf{A552} (2005) 566},
  \href{http://arxiv.org/abs/physics/0503191}{{\normalfont\ttfamily
  arXiv:physics/0503191}}\relax
\mciteBstWouldAddEndPuncttrue
\mciteSetBstMidEndSepPunct{\mcitedefaultmidpunct}
{\mcitedefaultendpunct}{\mcitedefaultseppunct}\relax
\EndOfBibitem
\bibitem{Hocker:2007ht}
H.~Voss, A.~Hoecker, J.~Stelzer, and F.~Tegenfeldt,
  \ifthenelse{\boolean{articletitles}}{\emph{{TMVA - Toolkit for Multivariate
  Data Analysis with ROOT}}, }{}\href{https://doi.org/10.22323/1.050.0040}{PoS
  \textbf{ACAT} (2007) 040}\relax
\mciteBstWouldAddEndPuncttrue
\mciteSetBstMidEndSepPunct{\mcitedefaultmidpunct}
{\mcitedefaultendpunct}{\mcitedefaultseppunct}\relax
\EndOfBibitem
\bibitem{Xie:2009rka}
Y.~Xie, \ifthenelse{\boolean{articletitles}}{\emph{{sFit: a method for
  background subtraction in maximum likelihood fit}},
  }{}\href{http://arxiv.org/abs/0905.0724}{{\normalfont\ttfamily
  arXiv:0905.0724}}\relax
\mciteBstWouldAddEndPuncttrue
\mciteSetBstMidEndSepPunct{\mcitedefaultmidpunct}
{\mcitedefaultendpunct}{\mcitedefaultseppunct}\relax
\EndOfBibitem
\bibitem{Cowan:2010js}
G.~Cowan, K.~Cranmer, E.~Gross, and O.~Vitells,
  \ifthenelse{\boolean{articletitles}}{\emph{{Asymptotic formulae for
  likelihood-based tests of new physics}},
  }{}\href{https://doi.org/10.1140/epjc/s10052-011-1554-0}{Eur.\ Phys.\ J.\
  \textbf{C71} (2011) 1554},
  \href{http://arxiv.org/abs/1007.1727}{{\normalfont\ttfamily
  arXiv:1007.1727}}, [Erratum: Eur. Phys. J. \textbf{C73}, 2501 (2013)]\relax
\mciteBstWouldAddEndPuncttrue
\mciteSetBstMidEndSepPunct{\mcitedefaultmidpunct}
{\mcitedefaultendpunct}{\mcitedefaultseppunct}\relax
\EndOfBibitem
\bibitem{Skwarnicki:1986xj}
T.~Skwarnicki, {\em {A study of the radiative cascade transitions between the
  Upsilon-prime and Upsilon resonances}}, PhD thesis, Institute of Nuclear
  Physics, Krakow, 1986,
  {\href{http://inspirehep.net/record/230779/}{DESY-F31-86-02}}\relax
\mciteBstWouldAddEndPuncttrue
\mciteSetBstMidEndSepPunct{\mcitedefaultmidpunct}
{\mcitedefaultendpunct}{\mcitedefaultseppunct}\relax
\EndOfBibitem
\bibitem{LHCb-PAPER-2015-010}
LHCb collaboration, R.~Aaij {\em et~al.},
  \ifthenelse{\boolean{articletitles}}{\emph{{Observation of the decay
  \mbox{\decay{\Bsb}{\psitwos\Kp\pim}}}},
  }{}\href{https://doi.org/10.1016/j.physletb.2015.06.038}{Phys.\ Lett.\
  \textbf{B747} (2015) 484},
  \href{http://arxiv.org/abs/1503.07112}{{\normalfont\ttfamily
  arXiv:1503.07112}}\relax
\mciteBstWouldAddEndPuncttrue
\mciteSetBstMidEndSepPunct{\mcitedefaultmidpunct}
{\mcitedefaultendpunct}{\mcitedefaultseppunct}\relax
\EndOfBibitem
\bibitem{LHCb-PUB-2016-021}
L.~Anderlini {\em et~al.}, \ifthenelse{\boolean{articletitles}}{\emph{{The
  PIDCalib package}}, }{}
  \href{http://cdsweb.cern.ch/search?p=LHCb-PUB-2016-021&f=reportnumber&action_search=Search&c=LHCb+Notes}
  {LHCb-PUB-2016-021}, 2016\relax
\mciteBstWouldAddEndPuncttrue
\mciteSetBstMidEndSepPunct{\mcitedefaultmidpunct}
{\mcitedefaultendpunct}{\mcitedefaultseppunct}\relax
\EndOfBibitem
\bibitem{LHCb-PAPER-2017-021}
LHCb collaboration, R.~Aaij {\em et~al.},
  \ifthenelse{\boolean{articletitles}}{\emph{{Measurement of \CP observables in
  \mbox{\decay{\Bpm}{D^{(\ast)}\Kpm}} and \mbox{\decay{\Bpm}{D^{(\ast)}\pipm}}
  decays}}, }{}\href{https://doi.org/10.1016/j.physletb.2017.11.070}{Phys.\
  Lett.\  \textbf{B777} (2018) 16},
  \href{http://arxiv.org/abs/1708.06370}{{\normalfont\ttfamily
  arXiv:1708.06370}}\relax
\mciteBstWouldAddEndPuncttrue
\mciteSetBstMidEndSepPunct{\mcitedefaultmidpunct}
{\mcitedefaultendpunct}{\mcitedefaultseppunct}\relax
\EndOfBibitem
\bibitem{LHCb-PAPER-2020-046}
LHCb collaboration, R.~Aaij {\em et~al.},
  \ifthenelse{\boolean{articletitles}}{\emph{{Precise measurement of the
  $f_s/f_d$ ratio of fragmentation fractions and of $B^0_s$ decay branching
  fractions}}, }{}\href{https://doi.org/10.1103/PhysRevD.104.032005}{Phys.\
  Rev.\  \textbf{D104} (2021) 032005},
  \href{http://arxiv.org/abs/2103.06810}{{\normalfont\ttfamily
  arXiv:2103.06810}}\relax
\mciteBstWouldAddEndPuncttrue
\mciteSetBstMidEndSepPunct{\mcitedefaultmidpunct}
{\mcitedefaultendpunct}{\mcitedefaultseppunct}\relax
\EndOfBibitem
\bibitem{LHCb-PAPER-2018-014}
LHCb collaboration, R.~Aaij {\em et~al.},
  \ifthenelse{\boolean{articletitles}}{\emph{{Observation of the decay
  \mbox{\decay{\Bs}{\Dbar^0 \Kp\Km}}}},
  }{}\href{https://doi.org/10.1103/PhysRevD.98.072006}{Phys.\ Rev.\
  \textbf{D98} (2018) 072006},
  \href{http://arxiv.org/abs/1807.01891}{{\normalfont\ttfamily
  arXiv:1807.01891}}\relax
\mciteBstWouldAddEndPuncttrue
\mciteSetBstMidEndSepPunct{\mcitedefaultmidpunct}
{\mcitedefaultendpunct}{\mcitedefaultseppunct}\relax
\EndOfBibitem
\bibitem{Belle:2007kff}
Belle collaboration, V.~Balagura {\em et~al.},
  \ifthenelse{\boolean{articletitles}}{\emph{{Observation of $\decay{\Dsop}{\Dp
  \pim\Kp}$ and angular decomposition of $\decay{\Dsop}{\Dstarp\KS}$}},
  }{}\href{https://doi.org/10.1103/PhysRevD.77.032001}{Phys.\ Rev.\
  \textbf{D77} (2008) 032001},
  \href{http://arxiv.org/abs/0709.4184}{{\normalfont\ttfamily
  arXiv:0709.4184}}\relax
\mciteBstWouldAddEndPuncttrue
\mciteSetBstMidEndSepPunct{\mcitedefaultmidpunct}
{\mcitedefaultendpunct}{\mcitedefaultseppunct}\relax
\EndOfBibitem
\end{mcitethebibliography}

\newpage
\centerline
{\large\bf LHCb collaboration}
\begin
{flushleft}
\small
R.~Aaij$^{33}$\lhcborcid{0000-0003-0533-1952},
A.S.W.~Abdelmotteleb$^{52}$\lhcborcid{0000-0001-7905-0542},
C.~Abellan~Beteta$^{46}$,
F.~Abudin{\'e}n$^{52}$\lhcborcid{0000-0002-6737-3528},
T.~Ackernley$^{56}$\lhcborcid{0000-0002-5951-3498},
B.~Adeva$^{42}$\lhcborcid{0000-0001-9756-3712},
M.~Adinolfi$^{50}$\lhcborcid{0000-0002-1326-1264},
P.~Adlarson$^{78}$\lhcborcid{0000-0001-6280-3851},
H.~Afsharnia$^{10}$,
C.~Agapopoulou$^{44}$\lhcborcid{0000-0002-2368-0147}
% ...

{\footnotesize \it

% ...
$^{10}$Universit{\'e} Clermont Auvergne, CNRS/IN2P3, LPC, Clermont-Ferrand, France\\
% ...
$^{33}$Nikhef National Institute for Subatomic Physics, Amsterdam, Netherlands\\
% ...
$^{42}$Instituto Galego de F{\'\i}sica de Altas Enerx{\'\i}as (IGFAE), Universidade de Santiago de Compostela, Santiago de Compostela, Spain\\
% ...
$^{44}$European Organization for Nuclear Research (CERN), Geneva, Switzerland\\
% ...
$^{46}$Physik-Institut, Universit{\"a}t Z{\"u}rich, Z{\"u}rich, Switzerland\\
% ...
$^{50}$H.H. Wills Physics Laboratory, University of Bristol, Bristol, United Kingdom\\
% ...
$^{52}$Department of Physics, University of Warwick, Coventry, United Kingdom\\
% ...
$^{55}$School of Physics and Astronomy, University of Glasgow, Glasgow, United Kingdom\\
$^{56}$Oliver Lodge Laboratory, University of Liverpool, Liverpool, United Kingdom\\
% ...
$^{78}$Department of Physics and Astronomy, Uppsala University, Uppsala, Sweden, associated to $^{55}$\\
% ...
}
\end{flushleft}

\end{document}